\documentclass[preprint,11pt]{elsarticle}

\usepackage[colorlinks,bookmarksopen,bookmarksnumbered,citecolor=red,urlcolor=red]{hyperref} 

\usepackage[utf8]{inputenc}
\usepackage{amsmath}
\usepackage{amssymb}
\usepackage{amsthm}
\usepackage{mathrsfs}
\usepackage{amsfonts}
\usepackage{bbold}
\usepackage[a4paper,margin=2.5cm]{geometry}
\usepackage{graphicx}
\usepackage{xifthen}
\usepackage[usenames,dvipsnames]{xcolor}

\newcommand{\skipifemptyarg}[1]{\ifthenelse{\isempty{#1}}{}{\left[#1\right]}}
\newcommand{\skipifscalar}[1]{\ifthenelse{\isempty{#1}}{}{;#1}}

\newcommand{\bs}[1]{\boldsymbol{#1}}
\newcommand{\scal}[2]{\bigl(#1,#2\bigr)}
\newcommand{\bilf}[2]{a\ifthenelse{\isempty{#1}}{}{\bigl(#1,#2\bigr)}}
\newcommand{\bilfN}[2]{\tilde{a}_\VN\ifthenelse{\isempty{#1}}{}{\bigl(#1,#2\bigr)}}
\newcommand{\set}[1]{\mathbb{#1}}

\newcommand{\MB}[1]{\boldsymbol{\mat{#1}}}
\newcommand{\mat}[1]{\mathsf{#1}} 
\newcommand{\V}[1]{\bs{#1}}
\newcommand{\T}[1]{\bs{#1}}
\newcommand{\mac}[1]{^{(#1)}}

\newcommand{\puc}{\mathcal{Y}}
\newcommand{\pVN}{\meas{\VN}}

\newcommand{\ptVNr}{\meas{\tVNr}}

\newcommand{\per}{\#}
\newcommand{\eff}{{\mathrm{H}}}

\newcommand{\bound}{^{\mathrm{bound}}}

\newcommand{\imu}{\mathrm{i}}		
\newcommand{\cA}{c_A}
\newcommand{\CA}{C_A}

\newcommand{\Aeff}{\TA_{\eff}}

\newcommand{\AeffN}{\TA_{\eff,\VN}}

\newcommand{\del}{\ensuremath{\delta}}
\newcommand{\alp}{\ensuremath{\alpha}}

\newcommand{\xR}{\set{R}}
\newcommand{\xN}{\set{N}}
\newcommand{\xZ}{\set{Z}}

\newcommand{\xRd}{{\set{R}^{d}}}
\newcommand{\xCd}{\set{C}^{d}}
\newcommand{\xCdN}{\set{C}^{d\times\VN}}
\newcommand{\xRdd}{\set{R}^{d\times d}}
\newcommand{\xCdd}{\set{C}^{d\times d}}
\newcommand{\xRdN}{\xR^{d\times \VN}}

\newcommand{\xRdtNr}{{\xR^{d\times(\tVNr)}}}

\newcommand{\xRddspd}{\set{R}_{\mathrm{spd}}^{d\times d}}

\newcommand{\xhMN}{\bigl[\set{C}^{d\times\VN}\bigr]^2}

\newcommand{\xNd}{\set{N}^d}
\newcommand{\Zd}{\set{Z}^d}
\newcommand{\ZdmO}{\set{Z}^d \backslash \{\T{0}\}}

\newcommand{\ZtNrd}{\set{Z}^d_{\tVNr}}
\newcommand{\ZNd}{\set{Z}^d_{\VN}}
\newcommand{\ZMd}{\xZ^d_{\V{M}}}

\newcommand{\cE}{\mathscr{E}}

\newcommand{\cEN}{\cE_\VN}

\newcommand{\Lper}[2]{L^{#1}_\#(\puc\skipifscalar{#2})}

\newcommand{\FT}[1]{\widehat{#1}}
\newcommand{\bfun}[1]{\varphi_{#1}}

\newcommand{\xE}{\set{E}}
\newcommand{\xEN}{\set{E}_\VN}

\newcommand{\cT}{\mathscr{T}}

\newcommand{\cTNd}{\cT_\VN^d}

\newcommand{\TA}{\ensuremath{\T{A}}}

\newcommand{\mhG}{\ensuremath{\T{\hat{\Gamma}}}}

\newcommand{\VN}{\ensuremath{{\V{N}}}}
\newcommand{\tVNr}{\ensuremath{{2\VN-\V{1}}}}

\newcommand{\VM}{\ensuremath{{\V{M}}}}

\newcommand{\Ve}{\ensuremath{\V{e}}}

\newcommand{\tVe}{\ensuremath{\Ve}}

\newcommand{\gVe}{\ensuremath{\widetilde{\Ve}}}

\newcommand{\VE}{\ensuremath{\V{E}}}

\newcommand{\Vu}{\ensuremath{\V{u}}}
\newcommand{\Vv}{\ensuremath{\V{v}}}

\newcommand{\Vx}{\ensuremath{\V{x}}}

\newcommand{\Vm}{\ensuremath{{\V{m}}}}
\newcommand{\Vn}{\ensuremath{{\V{n}}}}
\newcommand{\Vk}{\ensuremath{{\V{k}}}}

\newcommand{\Vo}{\ensuremath{{\V{0}}}}

\newcommand{\MBe}{\ensuremath{\MB{e}}}

\newcommand{\MBE}{\MB{E}}

\newcommand{\gMBe}{\ensuremath{\MB{\widetilde{e}}}}

\newcommand{\MBA}{\ensuremath{\MB{A}}}
\newcommand{\MBP}{\ensuremath{\MB{P}}}
\newcommand{\MBq}{\ensuremath{\MB{q}}}

\newcommand{\MBtA}{\ensuremath{\MB{\widetilde{A}}}}

\newcommand{\MBC}{\ensuremath{\MB{C}}}

\newcommand{\MBI}{\ensuremath{\MB{I}}}

\newcommand{\MBu}{\ensuremath{\MB{u}}}
\newcommand{\MBv}{\ensuremath{\MB{v}}}

\newcommand{\MBhv}{\ensuremath{\MB{\FT{v}}}}
\newcommand{\MBx}{\ensuremath{\MB{x}}}

\newcommand{\MBG}{\MB{G}}
\newcommand{\MBhG}{\ensuremath{\MB{\widehat{G}}}}

\newcommand{\MBF}{\ensuremath{\MB{F}}}
\newcommand{\MBFi}{\ensuremath{\MB{F}^{-1}}}

\newcommand{\MBGN}[1]{\ensuremath{\MBG_{\VN}^{#1}}}

\newcommand{\MBGNM}[1]{\ensuremath{\MBG_{\VN,\VM}^{#1}}}

\newcommand{\MBhGNM}[1]{\ensuremath{\MBhG_{\VN,\VM}^{#1}}}

\newcommand{\IN}{\mathcal{I}_\VN}
\newcommand{\INi}{\mathcal{I}_\VN^{-1}}

\newcommand{\ID}[1]{\mathcal{I}_{#1}}

\newcommand{\ItNr}{\mathcal{I}_\tVNr}

\newcommand{\norm}[2]{\bigl\| #1 \bigr\|_{#2}}
\newcommand{\meas}[1]{|#1|}
\newcommand{\mean}[1]{\langle#1\rangle}
\newcommand{\D}[1]{\,{\mathrm d}#1}

\theoremstyle{plain}

\newtheorem{algorithm}{Algorithm}

\newcommand{\tAeffN}{\widetilde{\TA}_{\eff,\VN}}

\newcommand{\sublabel}[1]{\raisebox{7mm}{\small (#1)}}
\newcommand{\trn}{^\mathsf{T}}

\journal{Journal of Computational Physics}

\begin{document}

\begin{frontmatter}

\title{A comparative study on low-memory iterative solvers for FFT-based homogenization of periodic media}

\author[label2]{Nachiketa Mishra}\ead{nachiketa.mishra@fsv.cvut.cz}
\author[label1,label2]{Jaroslav Vond\v{r}ejc}\ead{vondrejc@gmail.com}
\author[label2]{Jan Zeman\corref{cor1}}\ead{zemanj@cml.fsv.cvut.cz}

\cortext[cor1]{Corresponding author}

\address[label2]{Department of Mechanics, Faculty of Civil Engineering, Czech Technical  University in Prague, Th\'{a}kurova~7, 166~29 Prague 6, Czech Republic.}
 
\address[label1]{Institute of Scientific Computing, Technische Universität Braunschweig, Hans-Sommer-Stra{\ss}e~65, 381~06 Braunschweig, Germany}
 
\begin{abstract}
In this paper, we assess the performance of four iterative algorithms for solving non-symmetric rank-deficient linear systems arising in the FFT-based homogenization of heterogeneous materials defined by digital images.
Our framework is based on the Fourier-Galerkin method with exact and approximate integrations that has recently been shown to generalize the Lippmann-Schwinger setting of the original work by Moulinec and Suquet from 1994.
It follows from this variational format that the ensuing system of linear equations can be solved by general-purpose iterative algorithms for symmetric positive-definite systems, such as the Richardson, the Conjugate gradient, and the Chebyshev algorithms, that are compared here to the Eyre-Milton scheme---the most efficient specialized method currently available.
Our numerical experiments, carried out for two-dimensional elliptic problems, reveal that the Conjugate gradient algorithm is the most efficient option, while the Eyre-Milton method performs comparably to the Chebyshev semi-iteration.
The Richardson algorithm, equivalent to the still widely used original Moulinec-Suquet solver, exhibits the slowest convergence.
Besides this, we hope that our study highlights the potential of the well-established techniques of numerical linear algebra to further increase the efficiency of FFT-based homogenization methods.
\end{abstract}

\begin{keyword}
Numerical homogenization \sep Fourier-Galerkin method \sep Fast Fourier Transform \sep guaranteed bounds \sep Richardson iteration \sep Conjugate gradient algorithm \sep Chebyshev semi-iterative method \sep Eyre-Milton scheme
\end{keyword}
\end{frontmatter}

\section{Introduction}

Various experimental and simulation techniques, such as serial
sectioning~\cite{uchic_automated_2011}, computed
tomography~\cite{maire_quantitative_2013}, statistical
reconstruction~\cite{ballani_reconstruction_2015}, or digital models
\cite{sonon_advanced_2015} are currently available to characterize microstructures of
heterogeneous materials in a degree of realism not possible before. When
combined with the tools of homogenization theories,
e.g.~\cite{Cioranescu1999Intro2Homog,Milton2002TC,Geers2010}, 
these advances have made it possible to establish the structure-property
relations of complex engineering materials across length scales ranging from micrometers to tens of centimeters. The scale transitions rely on the solution of the \emph{corrector problem} -- a boundary value problem defined on a representative cell of the material, typically involving periodic boundary conditions. Since the input data are provided in the form of pixel- or voxel-based geometries, the need therefore arises for efficient solvers that employ images as discretization grids. Although several finite element or finite difference solvers have been developed for this purpose (e.g.~\cite{Terada:1997:DIB,Garboczi:1998:FEFD,pahr_high-resolution_2009}) methods based on the Fast Fourier Transform~(FFT) generally offer the best computational efficiency, because of the regular grid, the simple shape of the computational domain, and the periodic boundary conditions.

In the field of computational micromechanics of materials, the first FFT-based homogenization solver was proposed by Moulinec and Suquet in
1994~\cite{Moulinec1994FFT} and more than twenty years later, it is still widely used because of its simplicity and computational speed. The crux of the method is to reformulate the corrector problem as an integral equation of the Lippmann-Schwinger type solved by fixed-point iterations, while taking advantage of the fact that the kernel action can be efficiently handled using FFT. Later extensions of the basic algorithm were driven by the need to (i)~\emph{accelerate its convergence} for high-contrast problems~\cite{Eyre1999FNS,Michel2000CMB,Michel2001CSL,Vinogradov2008AFFT,Monchiet2012polarization};
(ii)~to increase \emph{accuracy of local fields} by incorporating inclusion shapes~\cite{Bonnet2007}, modified kernels~\cite{Willot2013fourier,Willot2015},
or local smoothing of coefficients~\cite{gelebart_filtering_2015,kabel_use_2015}; and to (iii)~\emph{prove the convergence of approximate solutions} in the framework of spectral collocation
methods~\cite{ZeVoNoMa2010AFFTH,VoZeMa2014FFTH,Schneider2014convergence}, the Galerkin discretization of the non-classical Hashin-Shtrikman functionals with piecewise-constant approximation
spaces~\cite{Brisard2012FFT,Brisard2014}, and standard Fourier-Galerkin methods~\cite{VoZeMa2014FFTH}.

Apart from providing theoretical justification to the original scheme, the Fourier-Galerkin setting has also been found convenient from the numerical point of view. For instance, it has clarified the effects of numerical quadrature~\cite{Vondrejc2015FFTimproved}, and led to the development of fully explicit guaranteed error bounds
on homogenized properties based on a primal-dual variational approach~\cite{VoZeMa2014FFTH,VoZeMa2014GarBounds}, which were later shown to be more restrictive than the corresponding Hashin-Shtrikman bounds~\cite{Monchiet2015}. The purpose of this paper is to complement these studies by examining the performance of four low-memory iterative methods for solving linear systems associated with the Fourier-Galerkin discretizations.  Our comparison involves general-purpose short-recurrence solvers, namely the Richardson scheme~\cite{richardson:1911}, the Conjugate gradient method~\cite{hestenes:1952:MCG}, the Chebyshev semi-iteration~\cite{Lanczos1953}, together with the Eyre-Milton algorithm~\cite{Eyre1999FNS} -- the most efficient of the accelerated schemes developed specifically for FFT-based homogenization problems, according to the recent study~\cite{Moulinec2014comparison}.

\paragraph{Related work} Previous comparative studies on FFT-based
homogenization algorithms fall into two categories. The aim of the first group of works is to compare their results with finite element solvers for material-specific applications, such as particle-reinforced composites with elasto-plastic phases~\cite{Michel1999}, visco-plastic models of polycrystalline materials~\cite{Prakash2009,Liu2010comparison,robert_comparison_2015}, or transport processes and creep in concrete-like
materials~\cite{Dunant2013a,Bary2014373}. Results of these studies consistently reveal that FFT-based methods offer at
least an order-of-magnitude improvement in the computational time while
predicting very similar distributions of local fields. The second group of studies is dedicated to accelerated schemes, namely to benchmarking their computational performance for high-contrast problems~\cite{Moulinec2003CFFT} and to revealing that they can be derived from a common recurrence relation~\cite{Moulinec2014comparison}.

\paragraph{Contributions} Although considerable effort has been spent on benchmarking FFT-based algorithms, neither of the studies above addresses conventional iterative solvers for symmetric positive-definite systems, the applicability of which follows naturally from the Fourier-Galerkin setting~\cite{VoZeMa2012LNSC,VoZeMa2014FFTH}. We aim to fill this gap while utilizing the standard techniques and results of numerical linear algebra. In particular, we discuss in detail the (i)~eigenvalue distribution of the system matrix, (ii)~effects of numerical integration, and reduction in (iii)~algebraic errors and (iv)~guaranteed bounds on homogenized properties during iterations. To the best of our knowledge, this is the first study addressing such aspects for FFT-based homogenization solvers.

\paragraph{Limitations} Because our goal is to provide basic insight into the behavior of the different linear solvers for FFT-based homogenization, we restrict our attention to the two-dimensional scalar linear elliptic problems with isotropic phases, moderate contrasts in coefficients, and discretizations not exceeding~$\approx 3,000,000$ unknowns (corresponding to a $1,999 \times 1,999$ pixel image). We also do not provide details about the overall computational time,
since all simulations were performed with an experimental Python-based code FFTHomPy, available at \url{https://github.com/vondrejc/FFTHomPy}, that is not optimized for speed. However, because our observations are based on well-established results of numerical linear algebra, they extend directly to more involved applications of FFT-based homogenization solvers reported in the literature, as evidenced by recent contributions~\cite{Shanthraj:2015:NRS,deGeus:2016:FSFFT,Zeman:2016:FEP}.

\paragraph{Organization of the paper} The remainder of the manuscript is organized as follows. The essentials of the Fourier-Galerkin discretization of the periodic corrector problem are briefly reviewed in Section~\ref{sec:background} following our more detailed expositions~\cite{VoZeMa2014FFTH,VoZeMa2014GarBounds,Vondrejc2015FFTimproved}. In Section~\ref{sec:Lin_solver}, we provide details for the linear iterative solvers considered in this study. Results of the numerical experiments are gathered in Section~\ref{sec:examples}, and the paper is concluded with the summary of the most important findings in Section~\ref{sec:conclusion}.

\paragraph{Notation} We will denote $d$-dimensional vectors and matrices by boldface letters, e.g. $\V{a} = \left(a_\alpha \right)_{\alpha=1,\ldots,d}\in \xRd$ or $\T{A} = (A_{\alpha\beta})_{\alpha,\beta=1,\ldots,d} \in \xRdd$. The Euclidean inner product will be referred to as $\scal{\bullet}{\bullet}_{\xRd}$ and the corresponding norm as $\| \bullet \|_{\xRd}$. By $\xRddspd$,  we will refer to the space of symmetric positive-definite $d \times d$ matrices. 

Vectors and matrices arising from discretization on regular grids will be denoted by the bold serif font in order to highlight their special structures. In particular, for a parameter $\VN \in \xNd$ related to the discretization along each coordinate and an index set $\ZNd$ enumerating the degrees of freedom, see ahead to~\eqref{eq:grid} for the exact specification, we use
\begin{align*}
\MB{a}_{\VN}
=
\left(
a_\alpha^\Vk
\right)_{\alpha=1,\ldots,d}^{\Vk \in \ZNd}
\in \xRdN,
&&
\MB{A}_{\VN}
=
\left(
A_{\alpha\beta}^{\Vk\Vm}
\right)_{\alpha,\beta=1,\ldots,d}^{\Vk,\Vm \in \ZNd}
\in [\xRdN]^2.
\end{align*}

The corresponding matrix-vector and matrix-matrix multiplication are understood as 
\begin{align*}
(\MB{A}_{\VN}\MB{a}_{\VN})^{\Vk} 
= 
\sum_{\Vm \in \ZNd}
\T{A}^{\Vk\Vm}\V{a}^{\Vm},
&&
(\MB{A}_{\VN} \MB{B}_{\VN})^{\Vk\Vm}
=
\sum_{\Vn \in \ZNd}
\T{A}^{\Vk\Vn}\T{B}^{\Vn\Vm}
\text{ for }
\Vk,\Vm \in \ZNd,
\end{align*}
and the space $\xRdN$ is endowed with the following inner product and norm
\begin{align*}
\scal{\MB{a}_{\VN}}{\MB{b}_{\VN}}_{\xRdN}
=
\frac{1}{\pVN}
\sum_{\Vk \in \ZNd}
\scal{\V{a}^\Vk}{\V{b}^\Vk}_{\xRd},
&&
\| \MB{a}_{\VN} \|^2_{\xRdN} = \scal{\MB{a}_{\VN}}{\MB{a}_{\VN}}_{\xRdN},
\end{align*}
where $\pVN = \prod_{\alp=1}^{d} N_\alp$. The same nomenclature is used for complex-valued quantities.

\section{Background}\label{sec:background}
The periodic corrector problem amounts to finding the matrix $\Aeff \in \xRddspd$ defined implicitly by the variational statement, e.g.~\cite[Chapter 13]{Milton2002TC},
\begin{align}\label{eq:homog_form}
\scal{\Aeff \VE}{\VE}_{\xRd} &= \min_{\Ve\in\cE} \bilf{\VE+\Ve}{\VE+\Ve}=\bilf{\VE+\tVe\mac{\VE}}{\VE+\tVe\mac{\VE}},
\end{align}
where $\VE$ is an arbitrary vector in $\xRd$. The bilinear form 
\begin{align}\label{eq:bilinear_forms}
\bilf{\Vu}{\Vv} 
= \int_\puc \scal{\TA(\Vx)\Vu(\Vx)}{\Vv(\Vx)}_{\xRd}\D{\Vx},
\end{align}
is defined on the space of the square-integrable $\xRd$-valued periodic functions on the unit cell $\puc = \prod_{\alp=1}^d \bigl(-\frac{1}{2}, \frac{1}{2}\bigr)$, denoted as $\Lper{2}{\xRd}$, and involves the matrix-valued coefficients $\TA: \puc \rightarrow \xRddspd$ that satisfy 
\begin{align}\label{eq:A}
\cA \norm{\Vv}{\xRd}^2
\leq 
\scal{\TA(\Vx)\Vv}{\Vv}_{\xRd}
\leq \CA \norm{\Vv}{\xRd}^2
\text{ for all } \Vv \in \xRd \text{ and almost all } \Vx \in \puc,
\end{align}
with  $0 < \cA \leq \CA < +\infty$.

The minimization problem~\eqref{eq:homog_form} is constrained to a subspace of $\Lper{2}{\xRd}$ defined by
\begin{align*}
\cE &= \{\Vv\in \Lper{2}{\xRd}:\text{curl }\Vv = \Vo,\int_\puc \Vv(\Vx) \D{\Vx} = \Vo\},
\end{align*}
which reflects the fact that admissible vectors can be expressed as the gradient of a $\puc$-periodic potential. 
This constraint can be conveniently enforced by the orthogonal projection operator $\mathcal{G}:\Lper{2}{\xRd} \rightarrow \cE$ given by, e.g.~\cite[Section 12.1]{Milton2002TC} or \cite[Lemma~2]{VoZeMa2014FFTH},
\begin{align}
\label{eq:proj}
\mathcal{G} [\Vv] &= 
\sum_{\Vk\in\Zd} \mhG(\Vk) \hat{\Vv}(\Vk) \varphi_{\Vk},
&
\mhG(\Vk) &=
\begin{cases}
\Vo
&\text{for }\Vk = \Vo,
\\
\frac{\Vk\Vk\trn}{\scal{\Vk}{\Vk}_{\xRd}}
&\text{for }\Vk\in\ZdmO,
\end{cases}
\end{align}
where $\mhG(\Vk)\in\xRdd$ are projection matrices in the Fourier space and $\FT{\Vv}(\Vk)\in\xCd$ is the $\Vk$-th Fourier coefficient of $\Vv$,
\begin{align}
\label{eq:FT_def}
\FT{\Vv}( \Vk )
&=
\int_\puc
\Vv(\Vx)
\bfun{-\Vk}(\Vx)
\D{\Vx}
\in \xCd
\quad\text{for }
\bfun{\Vk}(\Vx)
=
\exp
\Bigl(
  2\pi\imu
    \scal{\Vk}{\Vx}_{\xRd}
\Bigr)
\text{ with }\Vx \in \puc.
\end{align}

It follows from the Lax-Milgram lemma that~\eqref{eq:homog_form} has the unique minimizer $\tVe\mac{\VE}$ for any $\VE \in\xRd$ that satisfies the optimality conditions  
\begin{align*}
\bilf{\tVe\mac{\VE}}{\Vv} = - \bilf{\VE}{\Vv} & 
\text{ for all } \Vv\in\cE.
\end{align*}
In the following sections, we will explain how to obtain computable approximation to $\tVe\mac{\VE}$ using Fourier-Galerkin methods. 
This procedure includes specification of the approximating functions, Section~\ref{sec:trig_poly}; Galerkin discretizations with approximate and exact integrations yielding the guaranteed upper bounds on the homogenized matrix $\Aeff$,  Section~\ref{sec:galerkin_approx}; and the specification of linear systems resulting from the discretization procedure, Section~\ref{sec:linear-systems}.   

\subsection{Trigonometric polynomials}\label{sec:trig_poly}
Given the order of the polynomial approximation
\begin{align*}
\VN&\in\xN^d \text{ such that } N_\alp \text{ is odd for all }\alp,
\end{align*}
the space of \emph{$\xRd$-valued trigonometric polynomials} admits two equivalent definitions, e.g.~\cite[Chapter~8]{SaVa2000PIaPDE},
\begin{align}
\label{eq:trig_pol_space}
\cTNd &= 
\Bigl\{\sum_{\Vk\in\ZNd}{\FT{\Vv}^\Vk\varphi_{ \Vk }} : \FT{\Vv}^\Vk = \overline{\FT{\Vv}^{-\Vk}}\in\xCd\Bigr\}=\Bigl\{\sum_{\Vm\in\ZNd}{\Vv^\Vm\varphi_{\VN,\Vm}}: \Vv^\Vm\in\xRd\Bigr\}.
\end{align}
These definitions involve the set of truncated frequencies 
\begin{align}\label{eq:grid}
\ZNd  = 
  \left\{ \Vk \in \set{Z}^d : 
     |k_\alpha| < \frac{N_\alpha}{2} \right\},
\end{align}
the Fourier basis functions $\varphi_{\Vk}$ from~\eqref{eq:FT_def}, and the fundamental trigonometric polynomials
\begin{align*}
\varphi_{\VN,\Vm}(\Vx)
=
\frac{1}{\pVN}
\sum_{\Vk \in \ZNd}
\omega_{\VN}^{-\Vm\Vk} \varphi_{\Vk}(\Vx)
\text{ for }
\Vx \in \puc,
\end{align*}
where the complex-valued coefficients  
\begin{align*}
\omega_{\VN}^{\Vm\Vk} &=
\exp   \left(2 \pi \imu\sum_{\alp} \frac{m_\alp k_\alp}{N_\alp}  \right)
\text{ for }\Vm,\Vk\in \ZNd
\end{align*}
define the discrete Fourier transform~(DFT). 

Notice that the components $\FT{\Vv}^\Vk$ and $\Vv^\Vm$ in~\eqref{eq:trig_pol_space} associated with a trigonometric polynomial $\Vu_{\VN} \in \cTNd$ are not independent since they satisfy 
\begin{align*}
\FT{\Vv}^\Vk = \FT{\Vu}_\VN(\Vk), \quad
\Vv^\Vm = \Vu_{\VN} (\Vx_\VN^\Vm), \quad
\text{ for } 
\Vx_\VN^\Vm = \left( \frac{m_\alp}{N_\alp} \right)_{\alp=1,\ldots,d}
\text{ and } \Vk, \Vm \in \ZNd,
\end{align*}
where $\{ \Vx_\VN^\Vm \}^{\Vm \in \ZNd}$ denotes the regular grid in the real space $\xRd$ associated with the polynomial order $\VN$. Thus, the matrices $\MBhv_{\VN} = (\FT{\Vv}^\Vk)^{\Vk \in \ZNd}$ and $\MBv_{\VN} = (\Vv^\Vm)^{\Vm \in \ZNd}$ can be mapped on each other with the help of DFT
\begin{align*}
\MBhv_{\VN} = \MBF_{\VN}\MBv_{\VN}\in\xCdN,
&&
\MBv_{\VN} = \MBFi_{\VN}\MBhv_{\VN}\in\xRdN,
\end{align*}
where the matrices implementing the forward and inverse DFT are given by
\begin{align}\label{eq:DFT}
\MBF_{\VN} = \frac{1}{\pVN} \bigl( \del_{\alp\beta} \omega_{\VN}^{-\Vm\Vk} \bigr)_{\alp,\beta=1,\dotsc,d}^{\Vm,\Vk\in\ZNd} \in\xhMN ,
&&
\MBFi_{\VN} = \bigl( \del_{\alp\beta} \omega_{\VN}^{\Vm\Vk} \bigr)_{\alp,\beta=1,\dotsc,d}^{\Vm,\Vk\in\ZNd} \in\xhMN;
\end{align}
$\del_{\alp\beta}$ stands for the Kronecker delta equal to one for $\alp = \beta$, and to zero otherwise.

In the following sections, we shall make repeated use of the \emph{discretization operator} $\IN : C_{\per}(\puc;\xRd) \rightarrow \xRdN$,
\begin{align*}
\IN[\Vv] = \bigl( \Vv(\Vx_\VN^{\Vk}) \bigr)^{\Vk\in\ZNd},
\end{align*}
that assigns the values of a continuous periodic function $\Vv$ at the regular grid $\{ \Vx_{\VN}^\Vk \}^{\Vk \in \ZNd}$ to the vector from $\xRdN$. The operator $\IN$ establishes a scalar product-preserving one-to-one map between $\cTNd$ and $\xRdN$, with the inverse $\INi : \xRdN \rightarrow \cTNd$, 
and can thus be used with advantage to evaluate the action of the bilinear form~\eqref{eq:bilinear_forms} on trigonometric polynomials.

\subsection{Galerkin approximations and bounds}\label{sec:galerkin_approx}
The \emph{conforming} finite-dimensional space on which the Galerkin approximations will be performed consists of curl-free trigonometric polynomials with zero mean
\begin{align*}
\cEN = \mathcal{G}[\cTNd] = \cE \cap \cTNd.
\end{align*}
The homogenized matrix $\AeffN\in \xRddspd$, associated with the \emph{Galerkin approximation}~(Ga) to the corrector problem~\eqref{eq:homog_form}, then satisfies
\begin{align}\label{eq:Ga}
\scal{\AeffN \VE}{\VE}_{\xRd} &= \min_{\Ve_\VN\in\cEN} \bilf{\VE+\Ve_\VN}{\VE+\Ve_\VN} = \bilf{\VE+\Ve_\VN\mac{\VE}}{\VE+\Ve_\VN\mac{\VE}}
\end{align}
for arbitrary $\VE\in\xRd$. The most straightforward approach to the \emph{exact integration} in~\eqref{eq:Ga} utilizes the Plancherel theorem in the Fourier domain and leads to dense matrix representations, e.g.~\cite{Luciano1998} or~\cite[Section~6]{VoZeMa2014GarBounds}. However, sparsity is recovered when the integration is transferred to a double grid, as shown in~\cite[Section~6]{VoZeMa2014GarBounds}:
\begin{align}\label{eq:Ga_eval}
\bilf{\Vu_\VN}{\Vv_\VN} 
=
\scal{\MBA_\tVNr\MBu_{\VN,\tVNr}}{\MBv_{\VN,\tVNr}}_{\xRdtNr},
\end{align}
with $\MBu_{\VN,\tVNr} = \ItNr[\Vu_\VN]$, $\MBv_{\VN,\tVNr} = \ItNr[\Vv_\VN]$, and the block-diagonal matrix $\MBA_\tVNr$ provided by
\begin{align}\label{eq:Ga_exact_integration}
\MBA_{\tVNr}^{\Vk\Vm} =
\del^{\Vk\Vm} \sum_{\Vn\in\ZtNrd}
\omega_{\tVNr}^{\Vk\Vn}
\FT{\TA}(\Vn),
\end{align}
where $\FT{\TA}(\Vn) \in \xCdd$ denotes the $\Vn$-th Fourier coefficient of $\TA$, recall~\eqref{eq:FT_def}. For pixel- or voxel-wise constant coefficients, $\MBA_{\tVNr}$ can be assembled efficiently by FFT, see~\cite[Section~4]{Vondrejc2015FFTimproved} for details. 

Because the exact integration is relatively involved, we also introduce a simpler strategy based on the \emph{Galerkin approximation with numerical integration}~(GaNi). It employs the trapezoidal rule leading to discretization-dependent bilinear forms $\bilfN{}{} :\cTNd\times\cTNd\rightarrow\xR$,    
\begin{align*}
\bilf{\Vu_\VN}{\Vv_\VN}\approx\bilfN{\Vu_\VN}{\Vv_\VN} 
=
\frac{1}{\pVN}
\sum_{\Vk\in\ZNd}\TA(\Vx_\VN^\Vk)\Vu_\VN(\Vx_\VN^\Vk)\Vv_\VN(\Vx_\VN^\Vk) = \scal{\MBtA_\VN\MBu_\VN}{\MBv_\VN}_{\xRdN},
\end{align*}
where $\MBu_\VN=\IN[\Vu_\VN]$, $\MBv_\VN=\IN[\Vv_\VN]$, and the block-diagonal matrix $\MBtA_\VN = \bigl(\del^{\Vk\Vm} \TA (\Vx_\VN^\Vk)\bigr)^{\Vk,\Vm\in\ZNd}$ collects the coefficients at the grid points. In analogy to~\eqref{eq:homog_form} and~\eqref{eq:Ga}, the corresponding corrector problem amounts to finding the matrix $\tAeffN\in\xRddspd$ defined by
\begin{align}\label{eq:GaNi}
\scal{\tAeffN\VE}{\VE}_\xRd &= \min_{\Ve_{\VN}\in \cEN } \bilfN{\VE+\Ve_{\VN}}{\VE+\Ve_{\VN}} = \bilfN{\VE+\gVe_{\VN}\mac{\VE}}{\VE+\gVe_{\VN}\mac{\VE}}.
\end{align} 

Furthermore, combining the two conforming Galerkin schemes~\eqref{eq:Ga} and~\eqref{eq:GaNi} with the variational principle~\eqref{eq:homog_form}, one immediately obtains the \emph{guaranteed upper bounds} on the homogenized matrix 
\begin{align}\label{eq:bounds}
\scal{\Aeff\VE}{\VE}_{\xRd}
\leq 
\scal{\AeffN\VE}{\VE}_{\xRd}
\leq
\scal{\tAeffN\bound\VE}{\VE}_{\xRd}
=
\bilf{\VE+\gVe_\VN\mac{\VE}}{\VE+\gVe_\VN\mac{\VE}}
\quad\forall\VE\in\xRd,
\end{align}
where the matrix $\tAeffN\bound\in\xRddspd$ follows from the action of the bilinear form $a$ from~\eqref{eq:bilinear_forms} to the GaNi minimizers $\gVe_{\VN}\mac{\VE}$ evaluated with the exact integration formula~\eqref{eq:Ga_eval}, cf.~\cite[Section~6]{VoZeMa2014GarBounds} and~\cite[Section~4]{Vondrejc2015FFTimproved}. Note that analogous arguments establish guaranteed \emph{lower} bounds on $\Aeff$ according to the dual variational principle, finally leading to fully explicit \emph{discretization} error bounds on the approximate solutions. In the present paper, however, we shall work with the upper bounds~\eqref{eq:bounds} only; an interested reader is referred to~\cite{VoZeMa2014GarBounds,Vondrejc2015FFTimproved} for full details. 

\subsection{Linear systems}\label{sec:linear-systems}

The fully discrete versions of the optimality conditions for Ga~\eqref{eq:Ga} and GaNi~\eqref{eq:GaNi} follow from suitable applications of the discretization operators, resulting in 
\begin{subequations}\label{eq:fully_discrete_optimality}
\begin{align}
\scal{\MBA_\tVNr \MBe\mac{\VE}_{\VN,\tVNr}}{\MBv_{\VN,\tVNr}}_{\xRdtNr} &= -\scal{\MBA_{\VN,\tVNr}\MBE_\tVNr}{\MBv_\tVNr}_{\xRdtNr},
\\
\scal{\MBtA_\VN \gMBe\mac{\VE}_\VN}{\MBv_\VN}_{\xRdN} &= -\scal{\MBtA_\VN \MBE_\VN}{\MBv_\VN}_{\xRdN},
\end{align}
\end{subequations}
for all $\MBv_{\VN,\tVNr}\in\xE_{\VN,\tVNr}$ and $\MBv_\VN\in\xEN$, with
\begin{align*}
\xE_{\VN,\tVNr}
& = 
\ItNr[\cEN] \subset \xRdtNr,
&
\xEN &= \ID{\VN}[\cEN]\subset\xRdN,
\\
\MBe_{\VN,\tVNr}\mac{\VE} 
& = 
\ItNr[\Ve_\VN\mac{\VE}] \in\xE_{\VN,\tVNr},
&
\gMBe\mac{\VE}_{\VN} 
& = 
\ID{\VN}[\gVe_{\VN}\mac{\VE}]\in\xEN,
\\
\MBE_\tVNr & = \ItNr[\VE] \in \xRdtNr, 
&
\MBE_\VN & = \ID{\VN}[\VE] \in \xRdN,
\end{align*}
so, for example, $\xE_{\VN,\tVNr}$ contains the nodal values of trigonometric polynomials from $\cEN$ at the double grid points $\{ \Vx_\VN^\Vk \}^{\Vk \in \ZtNrd}$. 

To obtain the systems of linear equations defined by the optimality conditions~\eqref{eq:fully_discrete_optimality}, we need to enforce the constraints $\MBv_{\VN,\tVNr}\in\xE_{\VN,\tVNr}$ and $\MBv_\VN\in\xEN$ by means of suitable projections. Thanks to the properties of trigonometric polynomials, Section~\ref{sec:trig_poly}, such \emph{discrete orthogonal projections} follow directly from the continuous version~\eqref{eq:proj}: 
\begin{align}
\label{eq:discrete_proj}
\MBGNM{} &= \MBFi_\VM \MBhGNM{} \MBF_\VM
\quad\text{with }
\Bigl(\MBhGNM{}\Bigr)^{\Vk\Vm} = 
\begin{cases}
\Vo
&
\text{for }\Vk,\Vm\in\ZMd \backslash \ZNd
\\
\del^{\Vk\Vm}\mhG(\Vk),
&
\text{for }\Vk,\Vm\in\ZNd,
\\
\end{cases}
\end{align}
so that, e.g. $\MBG_{\VN,\tVNr} : \xRdtNr \rightarrow \xE_{\VN,\tVNr}$; we abbreviate $\MBG_{\VN,\VN}$ to $\MBGN{}$ in what follows. Now, the linear systems corresponding to Ga and GaNi arise as, cf.~\cite[Proposition~12]{VoZeMa2014FFTH} and \cite[Corollary~28]{Vondrejc2015FFTimproved},
\begin{subequations}
\label{eq:linsys}
\begin{align}
\label{eq:Ga_ls}
\MBG_{\VN,\tVNr} \MBA_\tVNr \MBe\mac{\VE}_{\VN,\tVNr} &= -\MBG_{\VN,\tVNr} \MBA_\tVNr \MBE_{\tVNr},
\\
\label{eq:GaNi_ls}
\MBGN{}\MBtA_\VN \gMBe\mac{\VE}_\VN &= -\MBGN{}\MBtA_\VN \MBE_{\VN}.
\end{align}
\end{subequations}

Once the solutions to these linear systems have been obtained, the guaranteed upper bounds from~\eqref{eq:bounds} can be made explicit:
\begin{subequations}
\label{eq:ev_bounds}
\begin{align}
\label{eq:ev_bounds_Ga}
\scal{\AeffN\VE}{\VE}_{\xRd} 
&= 
\scal{\MBA_\tVNr(\MBE_{\tVNr}+\MBe_{\VN,\tVNr}\mac{\VE})}{\MBE_{\tVNr}+\MBe_{\VN,\tVNr}\mac{\VE}}_{\xRdtNr},
\\
\label{eq:ev_bounds_GaNi}
\scal{\tAeffN\bound\VE}{\VE}_{\xRd} 
&= 
\scal{\MBA_\tVNr \mathcal{R}_{\VN,\tVNr}[\MBE_\VN + \gMBe_\VN\mac{\VE}]}{\mathcal{R}_{\VN,\tVNr}[\MBE_\VN + \gMBe_\VN\mac{\VE}]}_{\xRdtNr},
\end{align}
\end{subequations}
where the prolongation operator $\mathcal{R}_{\VN,\tVNr} = \ItNr \circ \INi$ maps $\xRdN$ to $\xRdtNr$ through the intermediate space of trigonometric polynomials $\cTNd$. 

\section{Solvers}
\label{sec:Lin_solver}

In order to avoid a profusion of notation, we shall refer to the linear systems~\eqref{eq:linsys} in a unified way as
\begin{align}\label{eq:linsy}
 \MB{C}\MB{x} = \MB{b}
 \text{ for }
 \MB{x} \in \xE,
\end{align}
so that, e.g., for the GaNi variant, $\MB{C} = \MBGN{}\MBtA_\VN$, $\MB{x} =
\gMBe\mac{\VE}_\VN$, $\MB{b} = -\MBGN{}\MBtA_\VN\MBE_\VN$, and $\xE = \xEN$. We shall also abbreviate $\scal{\bullet}{\bullet}_{\xRdN}$ or $\scal{\bullet}{\bullet}_{\xRdtNr}$ to $\scal{\bullet}{\bullet}_{2}$, $\norm{\bullet}{\xRdN}$ or $\norm{\bullet}{\xRdtNr}$ to $\norm{\bullet}{2}$, and $\MBE_{\VN}$ or $\MBE_{\tVNr}$ to $\MBE$ when there is no risk of confusion.

Notice that in general, matrix $\MB{C}$ in~\eqref{eq:linsy} is non-symmetric and highly rank-deficient. However, as first demonstrated by Vond\v{r}ejc et al.~\cite{VoZeMa2012LNSC}, $\MB{C}$ acts as a symmetric positive-definite matrix for vectors from the
subspace $\xE$ and satisfies
\begin{align*}
c_A \norm{\MB{x}}{2}^2
\leq 
\scal{\MB{C} \MB{x}}{\MB{x}}_{2}
\leq
C_A \norm{\MB{x}}{2}^2
\text{ for }
\MB{x} \in \xE,
\end{align*}
where $c_A$ and $C_A$ are the bounds on the material coefficients $\TA$ from~\eqref{eq:A}; the condition number of $\MBC$ on $\xE$ can be estimated from above \emph{independently of discretization} by $\kappa = C_A / c_A$ (we invite an interested reader to refer to Section~\ref{sec:eigen}, where
these properties are demonstrated on concrete examples). 

Because multiplication by $\MB{C}$ can be performed efficiently using the FFT in $\mathcal{O}(\meas{\VN}\log\meas{\VN})$ operations, recall the definition of projection operator in~\eqref{eq:discrete_proj} and~\eqref{eq:DFT}, the problem~\eqref{eq:linsy} turns out to be well-suited to conventional iterative algorithms for symmetric positive definite systems, provided that all iterates generated by the algorithm, $\MB{x}_m$, remain in $\xE$.

The remainder of this section is devoted to four such
iterative solvers, covering general-purpose short-recurrence iterative algorithms, Sections~\ref{sec:richard}--\ref{sec:cheby}; and
a special-purpose solver, Section~\ref{sec:milton}. In order to keep our exposition compact but self-contained, for each algorithm we present its pseudo-code; discuss the complexity of a single iteration, error estimates, and conformity of the iterates $\MBx_m \in \xE$; and briefly comment on their applications to Ga- and GaNi-based formulations in earlier studies.

\subsection{Richardson iteration}\label{sec:richard}
The Richardson iteration~\cite{richardson:1911} belongs to the group of stationary iterative methods. It searches for a fixed point of the mapping 
\begin{align}\label{eq:richard_iter}
\MB{x}_{m} 
= 
\MB{x}_{m-1}
- 
\omega(\MB{C}\MB{x}_{m-1} - \MB{b})
\text{ for }
m \in \xN 
\text{ and }
\MBx_0\in\xE
\end{align}
by means of Algorithm~\ref{alg:richard}.

\begin{algorithm}[Richardson iteration]\label{alg:richard}
Input: $\MB{C}$, $\MB{b}$, $\MB{x}_0 = \MB{0}$, $\cA$, $\CA$, and $\epsilon$\\
$\omega \longleftarrow \frac{2}{\cA + \CA}$~(optimal choice)\\
{\bf For} $m = 1,\;2,\;\cdots$\\
\phantom{x}\hspace{3ex} $\MB{x}_m \longleftarrow \MB{x}_{m-1} - \omega \left(
\MB{C}\MB{x}_{m-1} - \MB{b}\right)$\\
\qquad {\bf until} $\|\MB{x}_{m} - \MB{x}_{m-1}\|_2 \leq \epsilon \norm{\MBE}{2}$\\
{\bf return} $\MB{x}_m$
\end{algorithm}

A current implementation, one iteration of the Richardson solver involves three vectors, $\MBx_{m}$, $\MB{x}_{m-1}$, and $\MB{b}$, and one FFT-based matrix-vector multiplication $\MB{C}\MB{x}_{m-1}$ that dominates its computational cost. The convergence analysis of the Richardson scheme is available, e.g.,
in~\cite{young54} or~\cite[Section~4.2]{Saad:2003:IMSL}, where it is shown that the optimal choice of the iteration parameter 
\begin{align}\label{eq:optimal_richard}
\omega
=
\frac{2}{\cA + \CA}
\end{align}
provides the error bound for the $m$-th iteration in the form
\begin{align}\label{eq:Richard_estimate}
\norm{\MB{x}_{m} - \MB{x}}{2}
\leq
\left(
\frac{\kappa - 1}{\kappa + 1}
\right)^m
\norm{\MB{x}_0 - \MB{x}}{2}
\text{ for }
m \in \xN_0.
\end{align}
To see that all iterates are conforming to $\xE$, we rewrite the iterations
from Algorithm~\ref{alg:richard} as
\begin{align*}
\MB{x}_{m} = \MB{x}_{m-1} - \omega \MB{r}_{m-1}
\text{ for }
m \in \xN
\text{ and }
\MBx_0 \in \xE,
\end{align*}
where $\MB{r}_{m-1} = \MB{C} \MB{x}_{m-1} - \MB{b}$ is the $(m-1)$-th residual vector which is located in $\xE$, because both the matrix $\MB{C}$ and vector $\MB{b}$ involve the discrete projection operator; recall the linear systems~\eqref{eq:linsys}.  Hence, the approximate solutions are conforming, $\MB{x}_m \in \xE$, provided that the initial guess satisfies $\MB{x}_0 \in \xE$.

Rather interestingly, the Richardson iteration applied to the GaNi system~\eqref{eq:GaNi_ls} yields the original variant of the Moulinec-Suquet scheme~\cite{Moulinec1994FFT}; the optimal choice of the iteration parameter~\eqref{eq:optimal_richard} then corresponds to the results of convergence studies reported in, e.g.~\cite{Eyre1999FNS,Michel2001CSL,Vinogradov2008AFFT}.\footnote{%
Indeed, in the current notation, iterations of the Moulinec-Suquet algorithm are defined by the recurrence $\MB{x}_{m} = \MB{x}_{m-1} - \MB{\Gamma}_0 \MB{A} ( \MB{E} + \MB{x}_{m-1})$, where $\MB{\Gamma}_0 = \MB{G}/c_0$ stands for the (matrix) Green operator of the so-called reference problem with coefficient $c_0 \MB{I}$. For the optimal choice $c_0 = (c_A + C_A)/2$, the two algorithms coincide.} Applicability for Ga-based discretization has been recently reported by Monchiet~\cite{Monchiet2015}.

\subsection{Conjugate gradient method}\label{sec:CG}

The conjugate gradient method \cite{hestenes:1952:MCG} constructs the iterates by projecting the system~\eqref{eq:linsys} on to
a sequence of Krylov subspaces generated by the initial residual vector,
\begin{align}\label{eq:Krylov}
\set{K}_m = 
\mathrm{span}
\left\{ \MB{r}_0, \MBC\MB{r}_0, \MBC^2\MB{r}_0, \cdots,
\MBC^{m-1}\MB{r}_0 \right\} \mbox{ for } m \leq \dim(\xE),
\end{align}
utilizing a coupled two-term recurrence defined by~Algorithm~\ref{alg:cg}.

\begin{algorithm}[Conjugate gradients] 
Input: $\MBC$, $\MB{b}$, $\MB{x}_{0}$, and $\epsilon$ 
\\
\phantom{x}\hspace{3ex}$\MB{r}_{0} \longleftarrow \MB{b} - \MBC\MB{x}_0$\\
\phantom{x}\hspace{3ex}$\MB{p}_0 \longleftarrow \MB{r}_{0}$\\
\phantom{x}\hspace{3ex}{\bf For} $m = 1,\;2,\;\cdots$\\
\phantom{x}\hspace{3ex}\qquad $\alpha_{m-1} \longleftarrow \scal{\MB{r}_{m-1}}{\MB{r}_{m-1}}_{2}/
\scal{\MBC\MB{p}_{m-1}}{\MB{p}_{m-1}}_{2}$
\\
\phantom{x}\hspace{3ex}\qquad 
$\MB{x}_{m} \longleftarrow \MB{x}_{m-1} +
\alpha_{m-1}\MB{p}_{m-1}$
\\ 
\phantom{x}\hspace{3ex}\qquad 
$\MB{r}_{m}
\longleftarrow \MB{r}_{m-1} - {\alpha_{m-1}\MBC \MB{p}_{m-1}} $
\\ 
\phantom{x}\hspace{3ex}\qquad 
$\beta_{m-1} 
\longleftarrow 
\scal{\MB{r}_{m}}{\MB{r}_{m}}_{2}/
\scal{\MB{r}_{m-1}}{\MB{r}_{m-1}}_{2}$
\\
\phantom{x}\hspace{3ex}\qquad $\MB{p}_{m} = \MB{r}_{m} + \beta_{m-1}\MB{p}_{m-1}$\\
\phantom{x}\hspace{3ex}{\bf until} ${\|\MB{r}_{m}\|}_2 \leq \epsilon
\|\MB{b}\|_2$\\ \phantom{x}\hspace{3ex}{\bf return} $\MB{x}_{m}$\\
\label{alg:cg}
\end{algorithm}

In terms of storage, the Conjugate gradient method must keep track of three vectors, namely the solution $\MBx_{m}$, the residual $\MB{r}_m$, and the search direction $\MB{p}_m$. In addition, the product $\MBC \MB{p}_{m-1}$ must be calculated with the help of the FFT and stored to an auxiliary vector in order to keep the number of matrix-vector multiplications limited to one. Finally, notice that two scalar products are needed per a single iteration. 

As shown in, e.g.~\cite{Daniel1967,concus_golub_75}, the error in the
$m$-th iteration is bounded by
\begin{align}
\label{eq:CG_estimate}
\norm{\MB{x} - \MB{x}_{m}}{\MBC}
\leq 
2 \left(
\frac{\sqrt{\kappa}-1}{\sqrt{\kappa}+1}\right)^m
\norm{\MB{x} - \MB{x}_{0}}{\MBC}.
\end{align}
Since each Krylov subspace from~\eqref{eq:Krylov} satisfies $\set{K}_m \subset \xE$, the iterates $\MB{x}_m$ are conforming, as first observed in~\cite{VoZeMa2012LNSC}.

To the best of our knowledge, the first heuristic applications of the conjugate gradient method to systems associated with GaNi discretization was reported independently by Brisard and Dormieux~\cite{Brisard2010FFT} and Zeman~et al.~\cite{ZeVoNoMa2010AFFTH}, and justified later by Brisard and Dormieux~\cite{Brisard2012FFT} and Vondřejc~et al.~\cite{VoZeMa2012LNSC,VoZeMa2014FFTH}. Performance of the solver for the Ga system has been recently studied by
Vond\v{r}ejc~\cite{Vondrejc2015FFTimproved}, but no comparison with other iterative solvers has been made to date. 

\subsection{Chebyshev semi-iteration}\label{sec:cheby}
The Chebyshev semi-iteration~\cite{Lanczos1953,golub:varga:1961} builds upon a generalization of the Richardson iterative formula~\eqref{eq:richard_iter}
\begin{align}\label{eq:cheby_ite}
\MB{x}_{m} = \MB{x}_{m-1} + \omega_m \MB{r}_{m-1}
\text{ for }
m \in \xN 
\text{ and }
\MBx_0 \in \xE
,
\end{align}
where $\omega_m$ is related to the roots of the Chebyshev polynomials of the first kind, shifted to the interval $[\cA,\CA]$ and suitably normalized. In our numerical experiments, $\omega_m$ is determined
indirectly from a composite two-term recurrence according to
Algorithm~\ref{alg:chebychev}, because this relation proved to be the most numerically stable from the variants available in~\cite{Gutknecht:2002:PC}.

\begin{algorithm}[Chebyshev iteration]
\hspace{1ex}Input: $\MBC$, $\MB{b}$, $\MB{x}_0$, $\cA$, $\CA$, and $\epsilon$\\
\phantom{x}\hspace{3ex}$\MB{r}_{0} \longleftarrow \MB{b} - \MBC\MB{x}_0$\\
\phantom{x}\hspace{3ex}$c \longleftarrow \frac{1}{2}(\CA - \cA)$\\
\phantom{x}\hspace{3ex}$d \longleftarrow \frac{1}{2}(\cA + \CA)$\\
\phantom{x}\hspace{3ex}$\alpha \longleftarrow \frac{1}{d}$\\
\phantom{x}\hspace{3ex}$\beta \longleftarrow -\frac{1}{2}{\left( \frac{c}{d} \right)}^2 $\\
\phantom{x}\hspace{3ex}$\MB{r}_{0} \longleftarrow \MB{b} - \MBC\MB{x}_0$\\
\phantom{x}\hspace{3ex}$\MB{p}_{0} \longleftarrow \MB{r}_{0}$\\
\phantom{x}\hspace{3ex}{\bf For} $m = 1,\;2,\;\cdots$ \\
\phantom{x}\hspace{3ex}\qquad {\bf if} $m > 1$\\
\phantom{x}\hspace{3ex}\qquad \quad $ \quad \alpha \longleftarrow {\left(d + \frac{\beta}{\alpha} \right)}^{-1}$\\
\phantom{x}\hspace{3ex}\qquad \qquad $\beta \longleftarrow -{\left( \frac{c\alpha}{2} \right)}^2$\\
\phantom{x}\hspace{3ex}\qquad $\MB{r}_{m} \longleftarrow \MB{r}_{m-1} - \alpha\MBC\MB{p}_{m-1} $\\
\phantom{x}\hspace{3ex}\qquad $\MB{x}_{m} \longleftarrow \MB{x}_{m-1} + \alpha\MB{p}_{m-1}$\\
\phantom{x}\hspace{3ex}\qquad $\MB{p}_{m} \longleftarrow \MB{r}_{m} - \beta\MB{p}_{m-1}$ \\
\phantom{x}\hspace{3ex}{\bf until} ${\|\MB{r}_{m}\|}_2 \leq \epsilon
\|\MB{b}\|_2$\\ \phantom{x}\hspace{3ex}{\bf return} $\MB{x}_{m}$\\
\label{alg:chebychev}
\end{algorithm}

Similarly as in the case of the Conjugate gradient algorithm, the Chebyshev semi-iteration updates the triplet $(\MBx_{m}, \MB{r}_m, \MB{p}_m)$ at every iteration and requires only one matrix-vector product. 

As discussed, e.g., in~\cite[Section 2]{strakos:cheb2014}, the Conjugate gradient estimate~\eqref{eq:CG_estimate} holds also for the Chebyshev method and actually better reflects its true convergence behavior. By~\eqref{eq:cheby_ite}, all approximations $\MB{x}_m$ are confined by $\xE$. Note that, to the best of our knowledge, this work represents the first application of the Chebyshev method to FFT-based homogenization.

\subsection{Eyre-Milton accelerated scheme}\label{sec:milton}

The basic idea of the Eyre-Milton algorithm~\cite{Eyre1999FNS} is to recast the system~\eqref{eq:linsys} into an equivalent form with better conditioning and to solve the modified system by the Richardson scheme for the unknown in the form $\MBx= \MBE_\VN+\gMBe_\VN\mac{\VE}$. The resulting scheme appears in Algorithm~\ref{alg:eyre-milton}.

\begin{algorithm}[Eyre-Milton accelerated scheme]\label{alg:eyre-milton}
Input: $\MBA$, $\MBhG$, $\MB{x}_0 = \MBE$, $\cA$, $\CA$, 
and $\epsilon$\\ 
\phantom{x}\hspace{3ex}$\omega \longleftarrow \sqrt{\cA \CA}$\quad (Optimal choice)\\
\phantom{x}\hspace{3ex}$\MBP \longleftarrow {(\MBA + \omega\MBI)}^{-1}
\left(\MB{F}^{-1} [\MBI - 2\MBhG] \MB{F}\right) {(\MBA - \omega\MBI)}$\\
\phantom{x}\hspace{3ex}$\MBq = - 2 \omega{(\MBA + \omega\MBI)}^{-1}\MB{x}_0$\\
\phantom{x}\hspace{3ex}{\bf For} $m = 1,\;2,\;\cdots$\\
\phantom{x}\hspace{6ex} $\MBx_{m} = \MBP \MBx_{m-1} + \MBq$ \\
\phantom{x}\hspace{3ex}{\bf until} $\|\MB{x}_{m} - \MB{x}_{m-1}\|_2 \leq
\epsilon \norm{\MBE}{2}$ \\ \phantom{x}\hspace{3ex}{\bf return} $\MB{x}_{m}$
\end{algorithm}

The storage requirements for the Eyre-Milton scheme are similar to those for the Richardson iteration and involve $\MB{x}_{m-1}$, $\MB{x}_m$, and $\MB{q}$. Multiplication with matrix $\MBP$ is more demanding than with $\MBC$, cf.~\eqref{eq:linsys}, but its cost is dominated again by the forward and inverse FFT, and thus has the same complexity of $\mathcal{O}(\pVN \log \pVN)$.

The relative error bound for $\MBx_m$~\cite{Vinogradov2008AFFT}
\begin{align}\label{eq:E-M_estimate}
\norm{\MB{x}_m - \MB{x}}{2}
\leq
\left(\frac{\sqrt{\kappa}-1}{\sqrt{\kappa}+1}\right)^m
\norm{\MB{x}_m - \MB{x}_0}{2},
\end{align}
is similar to the Conjugate gradient and the Chebyshev estimates~\eqref{eq:CG_estimate}. In addition, the iterates generated by this algorithm are not conforming in the sense that $(\MBx - \MBE) \not\in \xE$, see~\cite{Moulinec2014comparison}. 

According to a seminal study by Moulinec and da Silva~\cite{Moulinec2014comparison}, the method outperforms other accelerated schemes available in the literature, namely the augmented Lagrangian formulation by Michel~et al.~\cite{Michel2000CMB,Michel2001CSL} and the polarization-based method by Monchiet and Bonnet~\cite{Monchiet2012polarization} in the~GaNi setting. However, the effect of numerical integration is very pronounced for this algorithm; quite surprisingly, we found that the method \emph{does not converge} when matrices in Algorithm~\ref{alg:eyre-milton} correspond to Ga discretization. We attribute this behavior to the fact that coefficient matrix for GaNi, $\MBtA_\VN$, still satisfies the estimates for the continuous problem~\eqref{eq:A},
\begin{align*}
\cA \norm{\MBv_\VN}{\xRdN}^2
\leq 
\scal{\MBtA_\VN\MBv_\VN}{\MBv_\VN}_{\xRdN}
\leq 
\CA \norm{\MBv_\VN}{\xRdN}^2
\text{ for all } \MBv_\VN \in \xRdN.
\end{align*}
However, a similar condition no longer holds for $\MBA_\tVNr$, because the exact integration impacts its eigenvalue distribution; see Section~\ref{sec:eigen} for an explicit example. 

\section{Examples}\label{sec:examples}

\begin{figure}[h]
\centering
\begin{tabular}{cc}
\sublabel{a}\includegraphics[height=55mm]{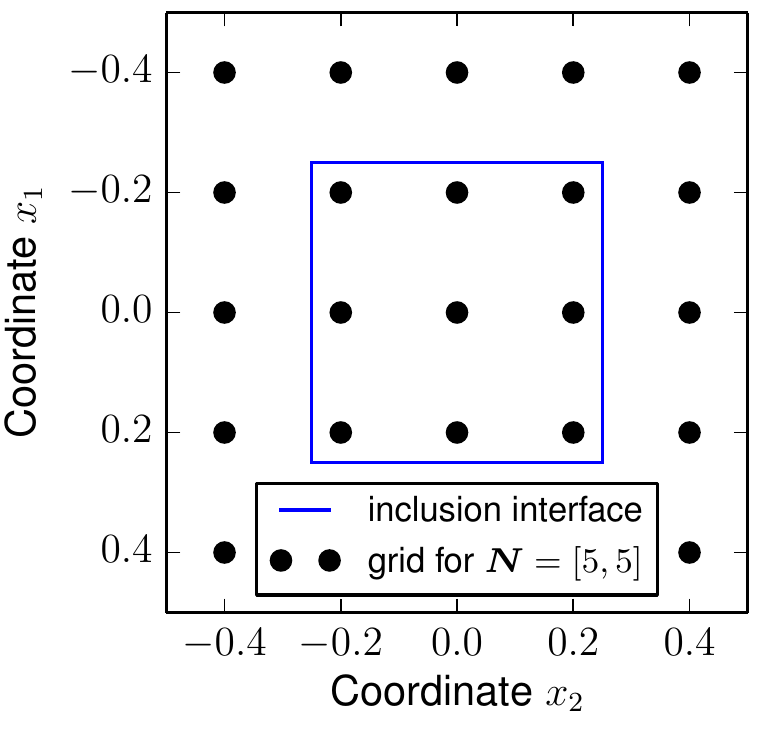}
& 
\sublabel{b}\includegraphics[height=55mm]{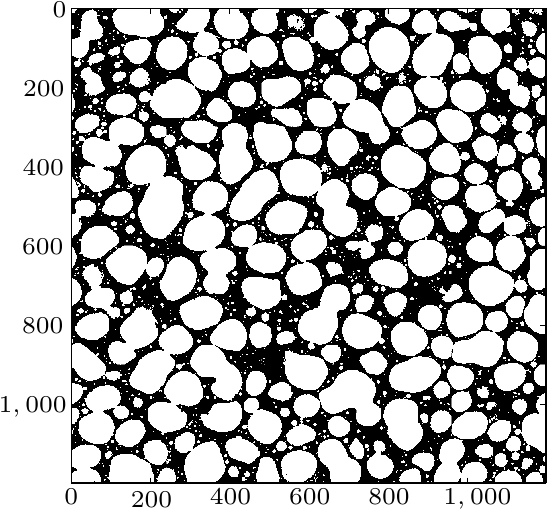}
\end{tabular}
\caption{(a)~An example of square inclusion discretized with $5\times 5$ points and (b)~$1,999 \times 1,999$ pixel bitmap of an alkali-activated fly-ash sample~(courtesy of Petr Hlaváček, CTU in Prague).} \label{fig:uc_types}
\end{figure}

Two types of unit cells appear in the comparative study. These
include a single square inclusion of $36\%$
volume fraction, Figure~\ref{fig:uc_types}(a), and a two-dimensional cross-section of an alkali-activated ash foam sample analyzed in~\cite{Hlavacek2014flyash}, Figure~\ref{fig:uc_types}(b). Unless specified otherwise, the coefficients
of the cell problem~\eqref{eq:A} are taken as
\begin{align}\label{eq:example_coefficients}
\TA = 
\begin{cases}
1 & \text{in matrix}, \\
100 & \text{in inclusion},
\end{cases}
&&
\TA = 
\begin{cases}
2.6 & \text{in matrix~(black)}, \\
0.026 & \text{in inclusion~(white)},
\end{cases}
\end{align}
for the square inclusion and the fly ash foam cell, respectively. The default
grid of the square inclusion is set to $85 \times 85$, since the same convergence behavior was observed for finer discretizations, while the alkali-activated foam sample corresponds to a $1,999 \times 1,999$ pixel bitmap. The macroscopic field is set to $\VE = [1;0]$ and the tolerance of all four algorithms is $\epsilon = 10^{-6}.$

\subsection{Eigenvalue distribution}\label{sec:eigen}

In order to obtain the complete eigenvalue distribution, the unit cells from
Figure~\ref{fig:uc_types} were discretized with trigonometric polynomials of order $15 \times 15$, leading to matrix sizes $d\pVN\times d\pVN=(2 \cdot 15 \cdot 15) \times (2 \cdot 15 \cdot 15) = 450
\times 450$ for GaNi discretization and $d\ptVNr\times d\ptVNr=(2 \cdot 29 \cdot 29) \times (2 \cdot
29 \cdot 29) = 1,682 \times 1,682$ for Ga. 

\begin{figure}[h]
\centering
\begin{tabular}{cc}
\multicolumn{2}{c}{\small \emph{Square inclusion}}\\
\sublabel{a}~\includegraphics[height=55mm]{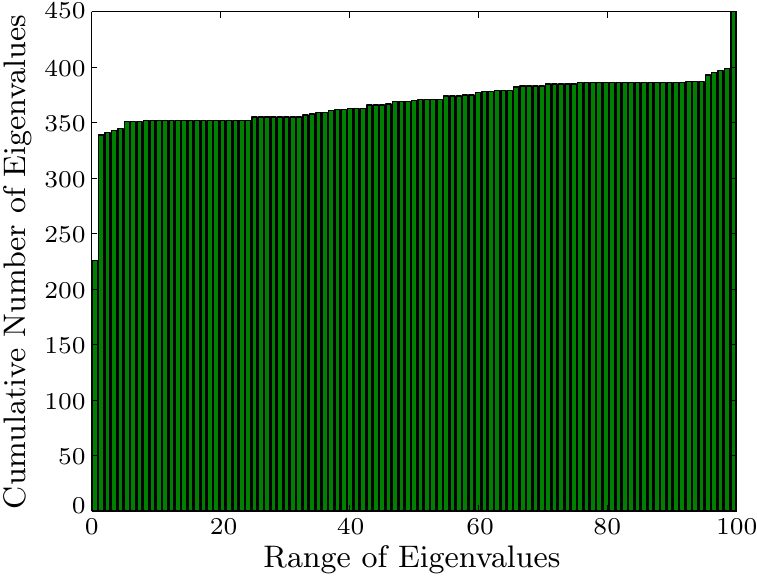} & 
\sublabel{b}~\includegraphics[height=55mm]{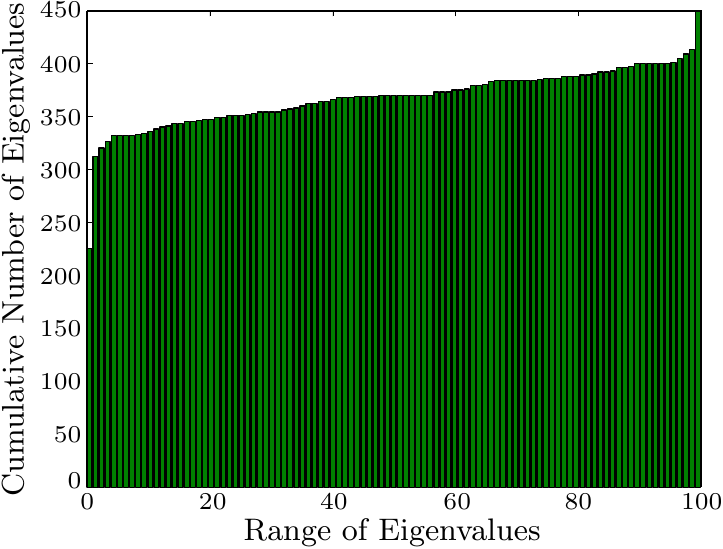} \\
\multicolumn{2}{c}{\small \emph{Fly ash foam}} 
\\
\sublabel{c}~\includegraphics[height=55mm]{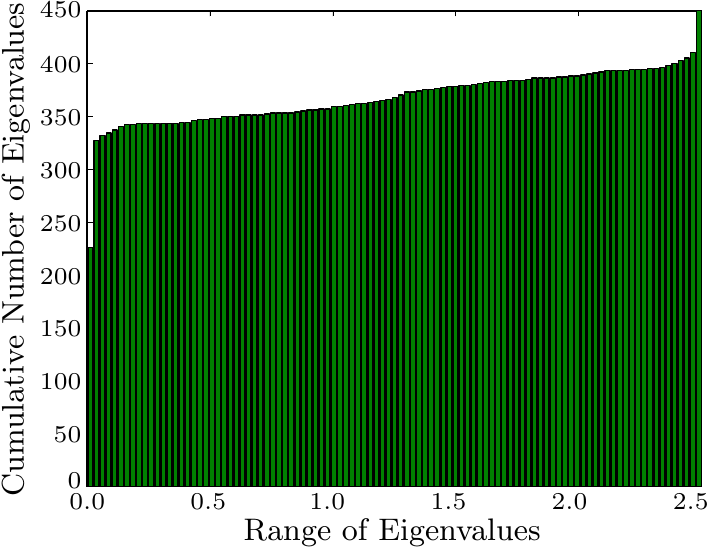} & 
\sublabel{d}~\includegraphics[height=55mm]{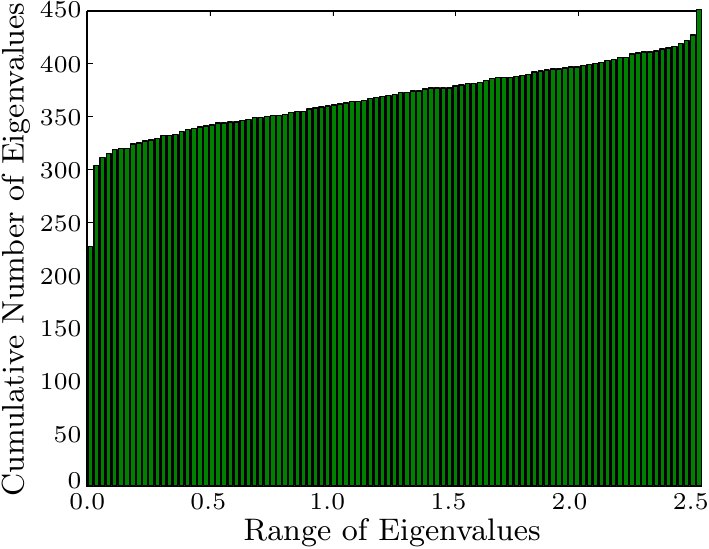}
\end{tabular}
\caption{Cumulative eigenvalue distribution for system matrices
resulting from GaNi~(left) and Ga~(right) discretizations; non-zero eigenvalues
are binned into $100$ intervals.}
\label{fig:eigenvalue_distrib}
\end{figure}

The cumulative distribution of eigenvalues for the two unit cell types and the
two discretizations appear in Figure~\ref{fig:eigenvalue_distrib}. For GaNi
discretization, the matrix has a rank of $224$, whereas the remaining $226$
eigenvalues are zero. It further follows from the discussion
in~\cite[Remark~30]{Vondrejc2015FFTimproved} that out of $226$ zero eigenvalues, $2$ correspond to the constant fields and $224$ represent trigonometric polynomials of zero divergence and zero mean. For Ga, the null-space is increased by $1,232$ eigenvectors resulting from the double-grid projection~\eqref{eq:discrete_proj}; notice that the corresponding eigenvalues are not included in Figure~\ref{fig:eigenvalue_distrib} for better clarity.

Figure~\ref{fig:eigenvalue_distrib} illustrates how cell geometry and numerical integration influence the matrix spectra. In particular, for the square inclusion and GaNi,
Figure~\ref{fig:eigenvalue_distrib}(a), the eigenvalues in spectrum $[1;100]$ form clusters. For instance no eigenvalues are present between $8$ and $25$ because the cumulative distribution is constant on this interval. Exact integration renders the spectrum less clustered, Figure~\ref{fig:eigenvalue_distrib}(b), but the effect of geometrical irregularity is even stronger, Figure~\ref{fig:eigenvalue_distrib}(c), so that the interval $[0.026;2.6]$ is sampled uniformly with the eigenvalues of the matrix resulting from full integration applied to the fly ash foam cell, Figure~\ref{fig:eigenvalue_distrib}(d). Finally, observe that irrespective of the discretization used, the non-zero eigenvalues are bounded by the coefficients of the phases according to~\eqref{eq:example_coefficients}; the matrix $\MBC$ is thus indeed symmetric positive-definite on $\xE$. The extreme
eigenvalues have the highest multiplicity as indicated by the corresponding jumps in the eigenvalue distribution.

\begin{figure}[h]
	\centering
	\begin{tabular}{cc}
		\sublabel{a}~\includegraphics[height=55mm]{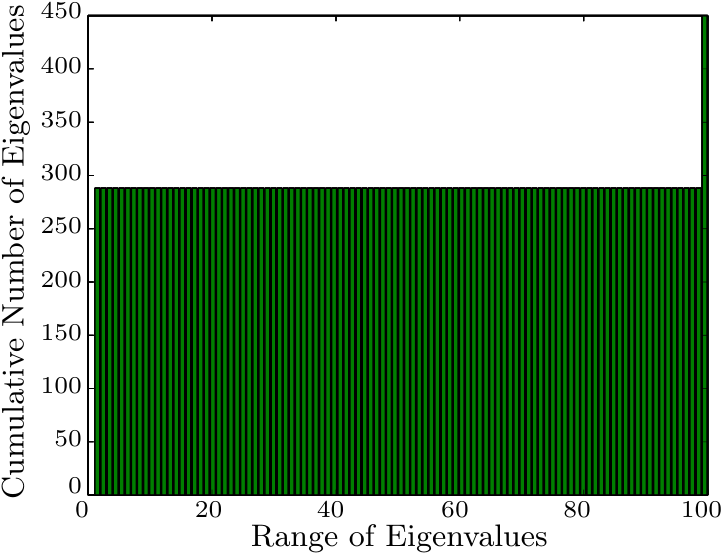} & 
		\sublabel{b}~\includegraphics[height=55mm]{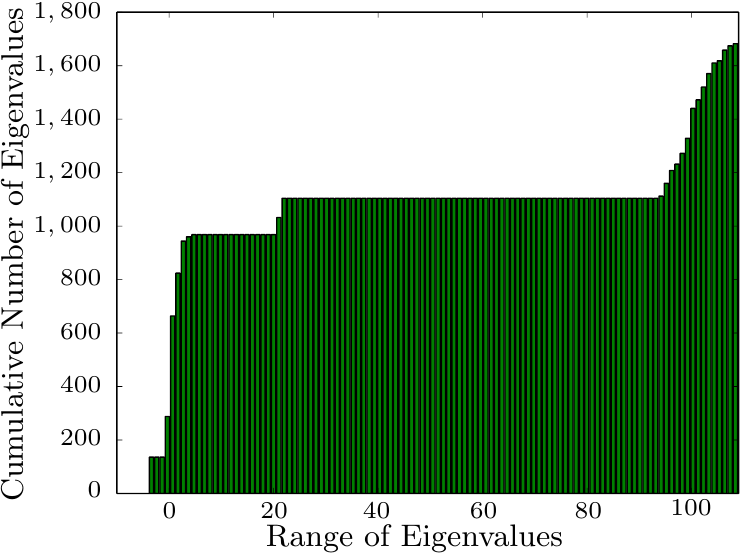} 
	\end{tabular}
	\caption{Cumulative eigenvalue distribution for coefficient matrices
		resulting from GaNi~(left) and Ga~(right) discretization of square inclusion; eigenvalues are binned into $100$ intervals.}
	\label{fig:eigenvalue_distrib_coefficients}
\end{figure}

Figure~\ref{fig:eigenvalue_distrib_coefficients} demonstrates the effect of numerical integration on matrices of coefficients $\MBtA_\VN$~(of rank 450) and $\MBA_\tVNr$~(of rank 1,682) for GaNi and Ga schemes and the square inclusion. In this case, the effects of integration are even more pronounced. For GaNi scheme, the spectrum consists of two values $\{1, 100\}$, because matrix $\MBtA_\VN$ contains only coefficients of the continuous problems~\eqref{eq:A} sampled on a regular grid. For Ga scheme, the exact integration~\eqref{eq:Ga_exact_integration} significantly changes the spectrum of $\MBA_\tVNr$; the eigenvalues are now located within interval $[-3.89,108.89]$. We conjecture that divergence of the Eyre-Milton scheme for Ga, reported in Section~\ref{sec:milton}, is a direct consequence of this fact.

\subsection{Residual norm}\label{sec:math_res_norm}

\begin{figure}[h]
\centering
\begin{tabular}{cc}
\multicolumn{2}{c}{\small \emph{Square inclusion}}\\
\sublabel{a}\includegraphics[height=55mm]{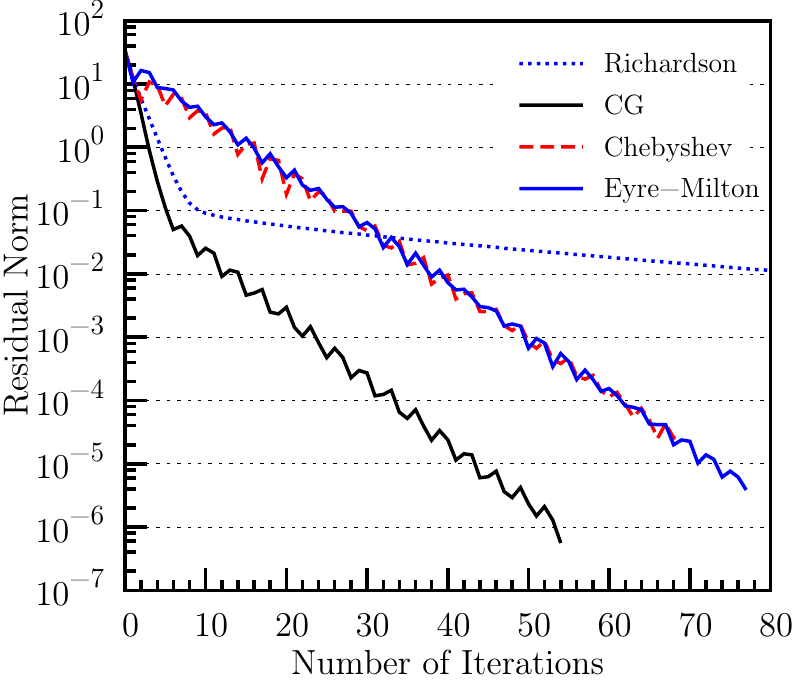} &
\sublabel{b}\includegraphics[height=55mm]{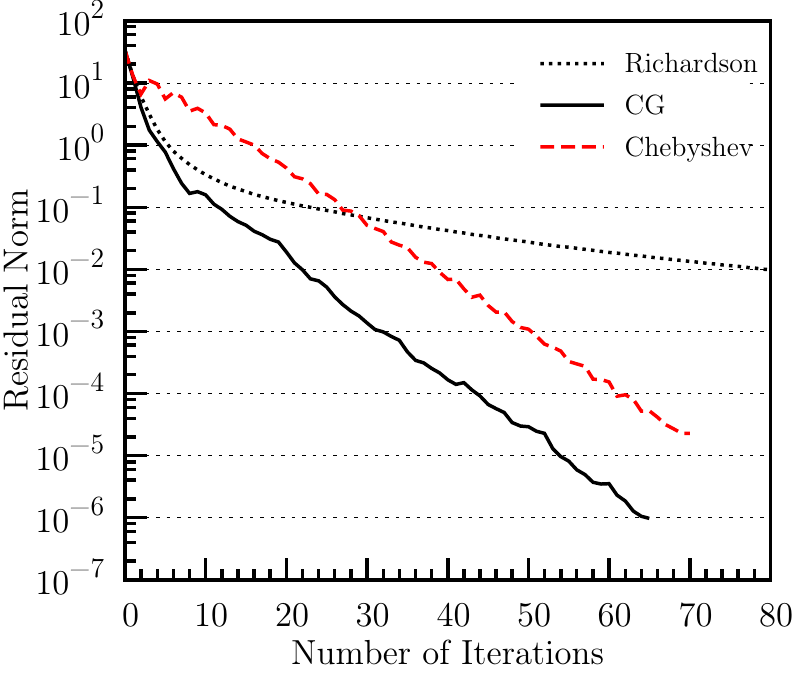}
\\[-2ex]
\multicolumn{2}{c}{\small \emph{Fly ash foam}} \\
\sublabel{c}\includegraphics[height=55mm]{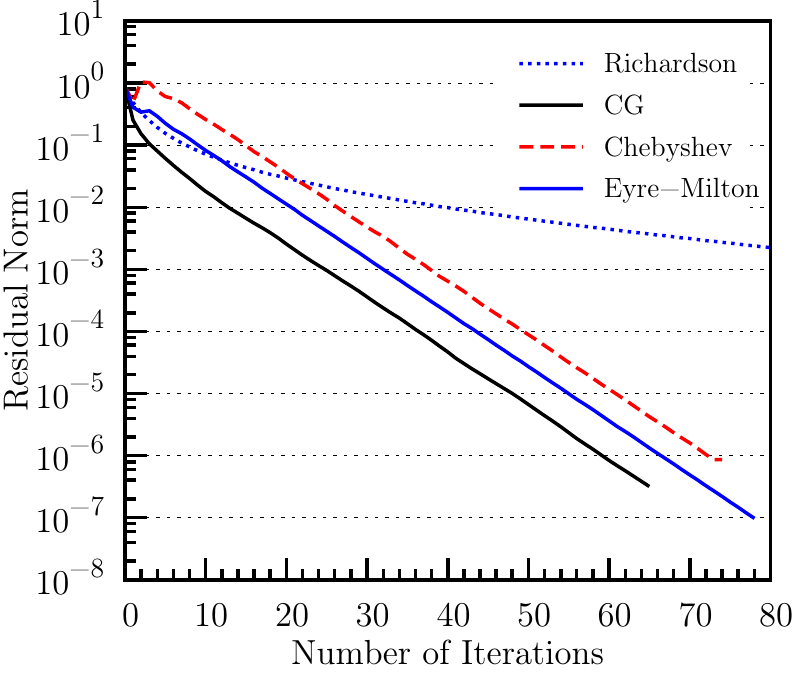} &
\sublabel{d}\includegraphics[height=55mm]{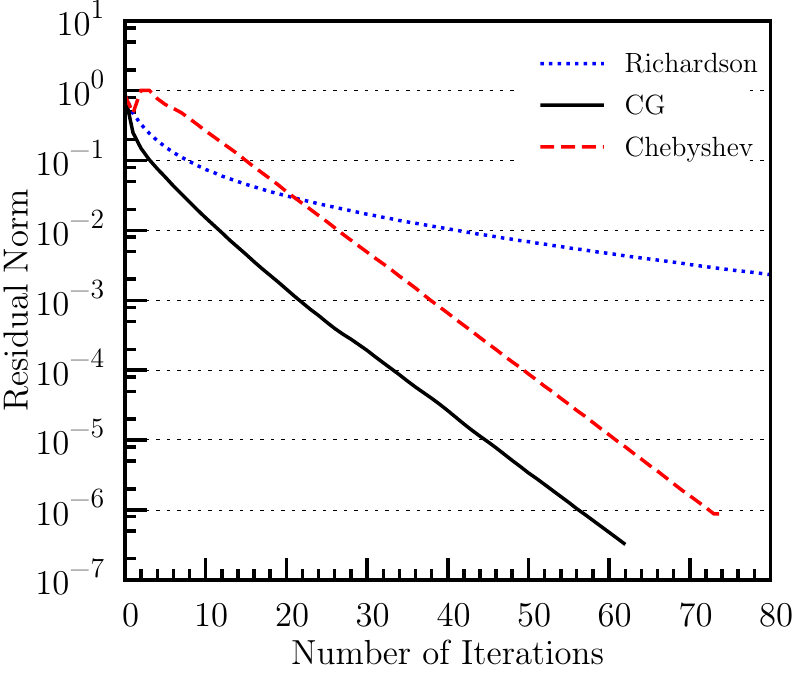}
\end{tabular}
\caption{Convergence history of residual norms for GaNi~(left) and Ga~(right) discretizations.}
\label{fig:convergence_residual}
\end{figure}

We start our study on the convergence properties of FFT-based solvers
by investigating the evolution of the residual norm
\begin{align*}
\norm{\MB{r}_m}{2}
=
\norm{\MB{b} - {\MBC}\MB{x}_m}{2}
\text{ for }
m \in \xN_0 
\end{align*}
during iterations, see Figure~\ref{fig:convergence_residual}. Results reveal that the Richardson scheme and Conjugate gradients display behavior qualitatively different from the Chebyshev method and the Eyre-Milton method. For the first two algorithms, convergence proceeds in two stages: the first stage~($\approx 10$ iterations) is associated with a rapid decrease in the residual norm, and then slows down in the second stage. These two stages are especially pronounced for the GaNi-discretized square inclusion, Figure~\ref{fig:convergence_residual}(a), but they are clearly visible in all remaining examples. 

The Chebyshev and the Eyre-Milton schemes display almost the same behavior along the whole iteration process. We attribute this behavior to the fact that the first group of algorithms initially resolves the
components of the residuum vectors associated with the largest eigenvalues and then proceeds through the rest of the spectrum down to the smallest eigenvalue, cf.~\cite{papez:strakos:2014}.
The second group of algorithms, on the other hand, simultaneously reduces the residuum components associated with the full spectrum, as confirmed by our computational observations that will form the basis of a separate publication.

Except for the Richardson scheme, all methods display oscillations in the convergence plots for regular geometry and GaNi, Figure~\ref{fig:convergence_residual}(a), that are significantly dampened by the exact integration and/or irregular distribution of phases. These phenomena again closely follow the eigenvalue distributions displayed in Figure~\ref{fig:eigenvalue_distrib}. We also
observe that the behavior of the Chebyshev method and Eyre-Milton scheme are almost identical for the simple geometry; for the fly ash foam microstructure, the Eyre-Milton method is more efficient and the residual norms generated by the Chebyshev method \emph{increase} in the first iteration. At later iterations, the rate of convergence of the Richardson scheme is inferior to the remaining algorithms, which exhibit super-linear convergence in accordance with error estimates~\eqref{eq:Richard_estimate}, \eqref{eq:CG_estimate}, and~\eqref{eq:E-M_estimate} reported earlier in
Section~\ref{sec:Lin_solver}. Nevertheless, the conjugate gradient method always displays the best performance.

\subsection{Guaranteed upper bound}\label{sec:guaranteed-upper-bound}

The numerical experiments in the previous section concentrated on the residual norm because it is used as the stopping criterion in conventional solvers. Here, we shall consider the evolution of the guaranteed upper bound at the $m$-th iteration,
\begin{align}\label{eq:guar_upper_bound_m}
U_{\VE}( \MBx_m )
& = 
\scal{\MBA_{\tVNr}( \MBE_{\tVNr} + \MB{x}_m)}{\MBE_{\tVNr} + 
\MBx_m}_{\xRdtNr} 
\text{ for }
m \in \xN_0,
\end{align}
where the mapping $U_{\VE}:\xE_{\VN,\tVNr}\rightarrow\xR$ provides the average energy in the unit cell with the distribution of local fields 
$(\VE + \mathcal{I}^{-1}_{\tVNr}[\MBx_m])$. If the iterates $\MBx_m\in\xE_{\VN,\tVNr}$ correspond to the Ga scheme, $U_{\VE}(\MBx_m) \longrightarrow \scal{\AeffN\VE}{\VE}_\xRd$, recall~\eqref{eq:ev_bounds_Ga}. For the GaNi scheme, the iterates $\MBx_m$ must be projected to $\xE_{\VN,\tVNr}$ to obtain $U_{\VE}(\mathcal{R}_{\VN,\tVNr}[\MBGN{}\MBx_m]) \longrightarrow \scal{\tAeffN\bound\VE}{\VE}_\xRd$, recall~\eqref{eq:ev_bounds_GaNi}. Notice that for $\MBx_0 = \MB{0}$, Eq.~\eqref{eq:guar_upper_bound_m} corresponds to the Voigt estimate.

\begin{figure}[h]
\centering
\begin{tabular}{cc}
\multicolumn{2}{c}{\small \emph{Square inclusion}}\\
\sublabel{a}\includegraphics[height=55mm]{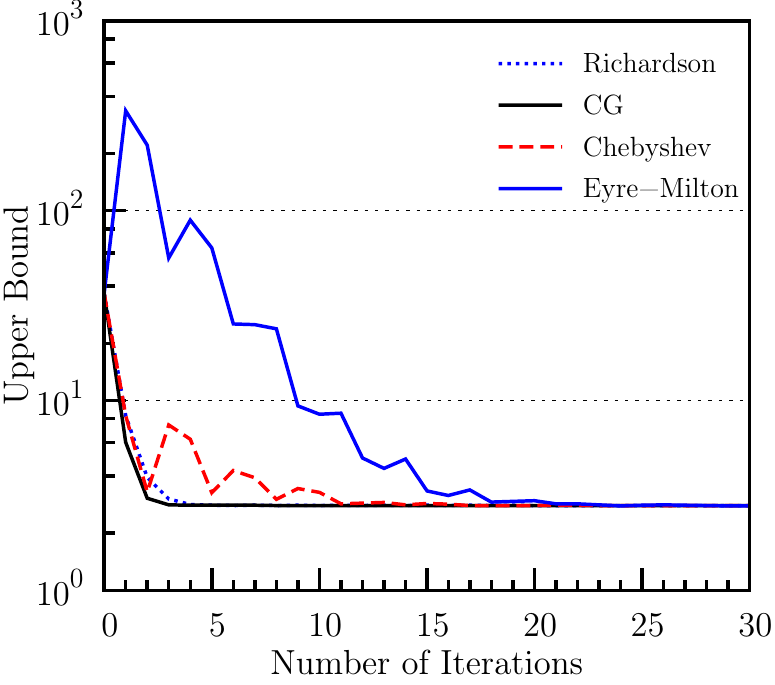}
&
\sublabel{b}\includegraphics[height=55mm]{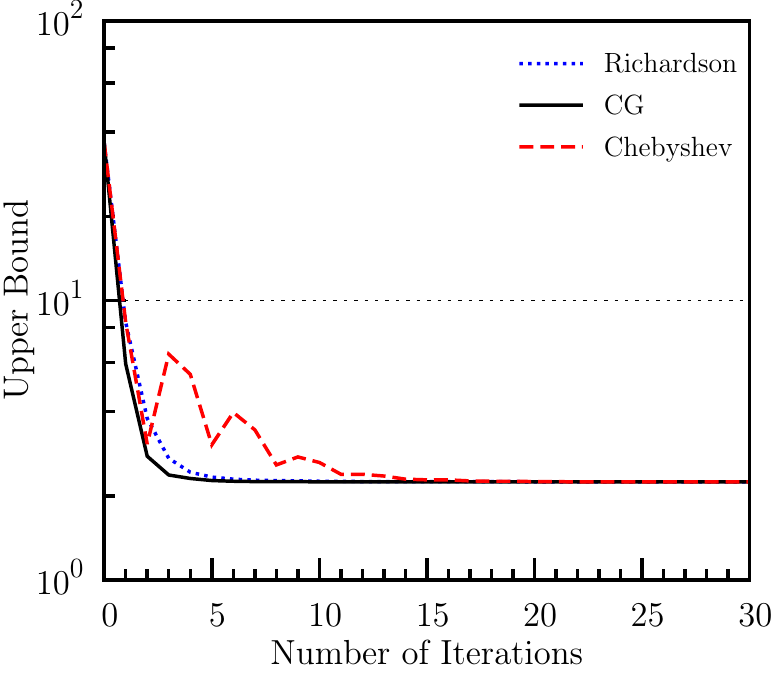}
\\[-2ex]
\multicolumn{2}{c}{\small \emph{Fly ash foam}} \\
\sublabel{c}\includegraphics[height=55mm]{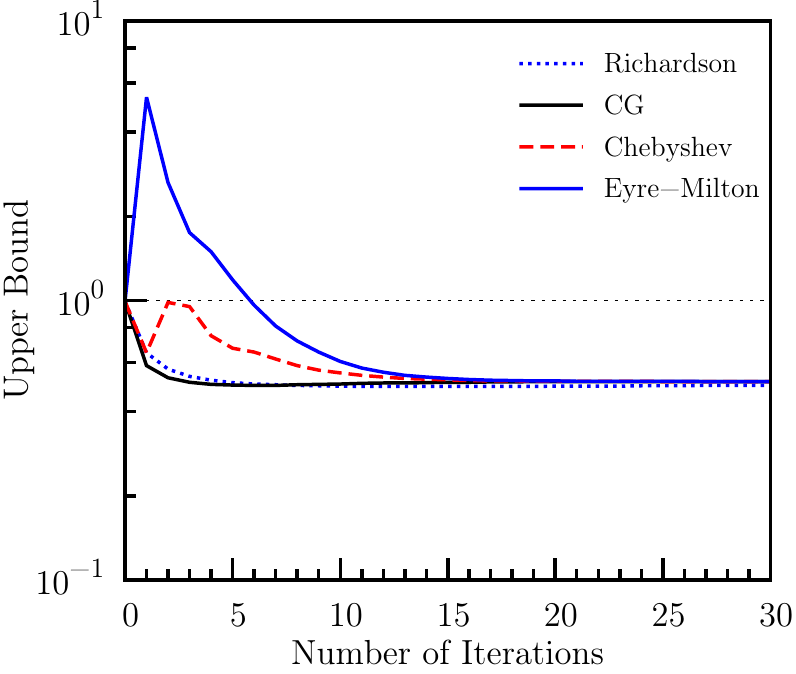}
&
\sublabel{d}\includegraphics[height=55mm]{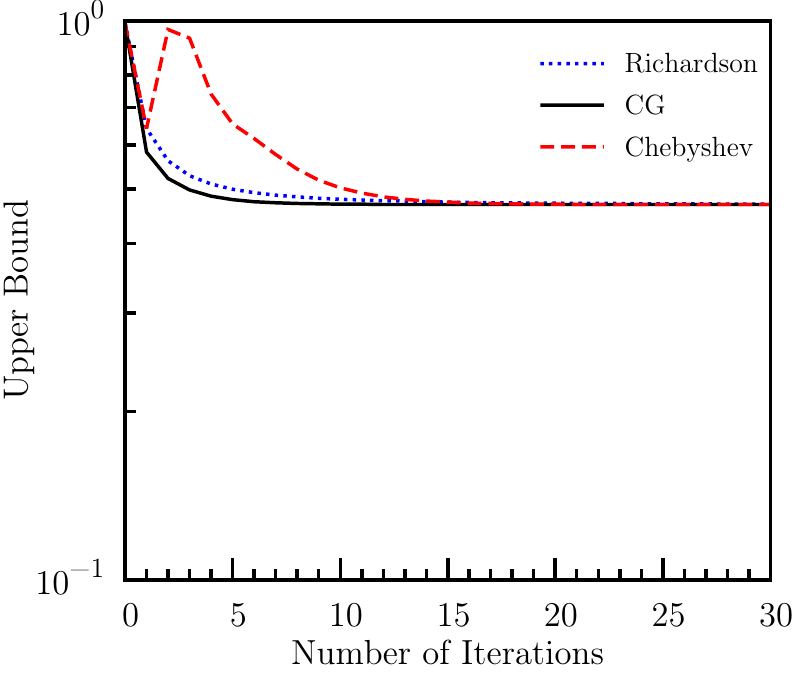}
\end{tabular}
\caption{Convergence history of guaranteed upper bounds for GaNi~(left) and
Ga~(right) discretizations.}
\label{fig:guar_bound_converge}
\end{figure}

The behavior of individual solvers, Figure~\ref{fig:guar_bound_converge}, agrees
well with the observations made from the residual plots,
Figure~\ref{fig:convergence_residual}; in particular the estimates on the guaranteed upper bound generated by the Richardson scheme and by the Conjugate gradient method converge faster than the Chebyshev and Eyre-Milton schemes in the first $\approx 10$ iteration, after which the bounds appear to stabilize. 

\begin{figure}[h]
\centering
\begin{tabular}{cc}
\multicolumn{2}{c}{\small \emph{Square inclusion}}\\
\sublabel{a}\includegraphics[height=55mm]{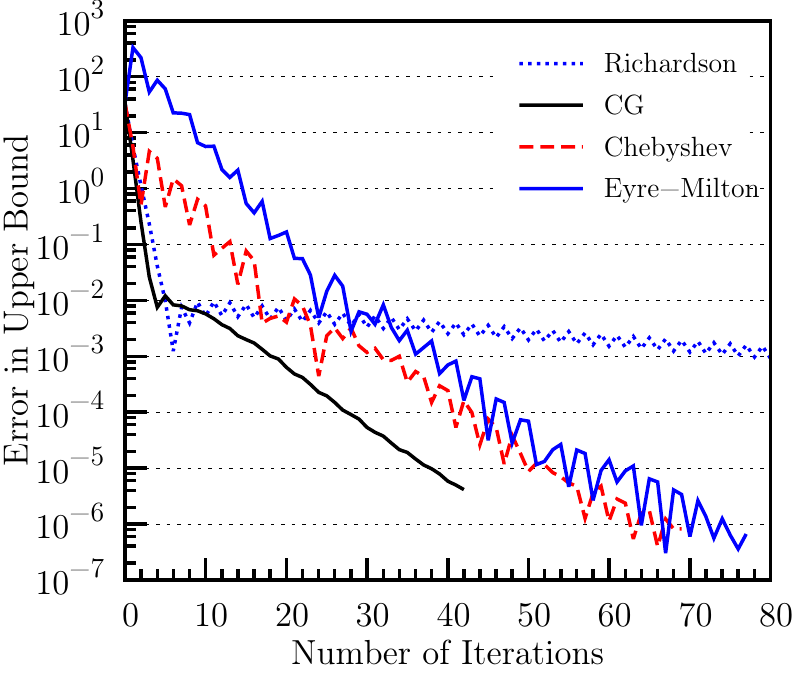}
&
\sublabel{b}\includegraphics[height=55mm]{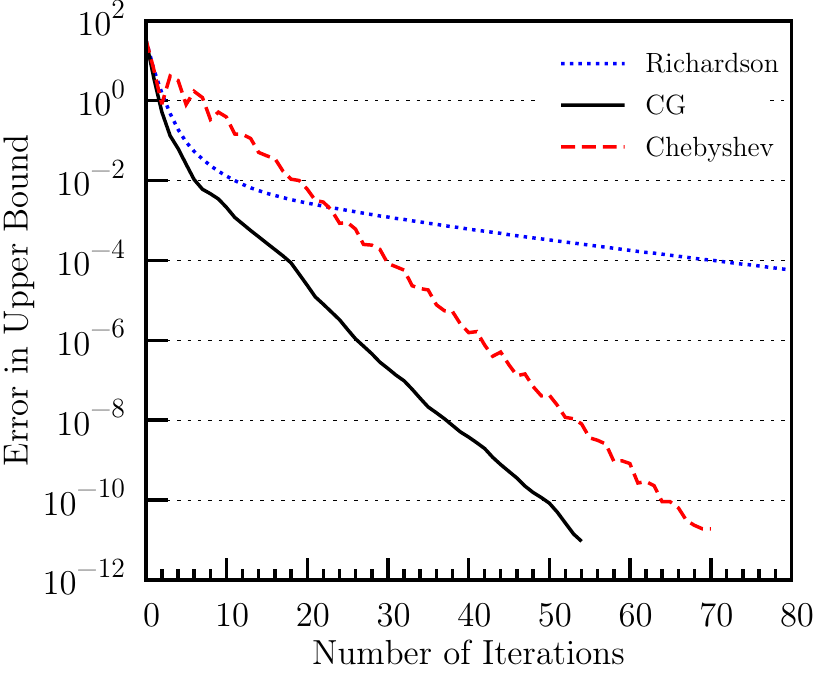}
\\[-2ex]
\multicolumn{2}{c}{\small \emph{Fly ash foam}} \\
\sublabel{c}\includegraphics[height=55mm]{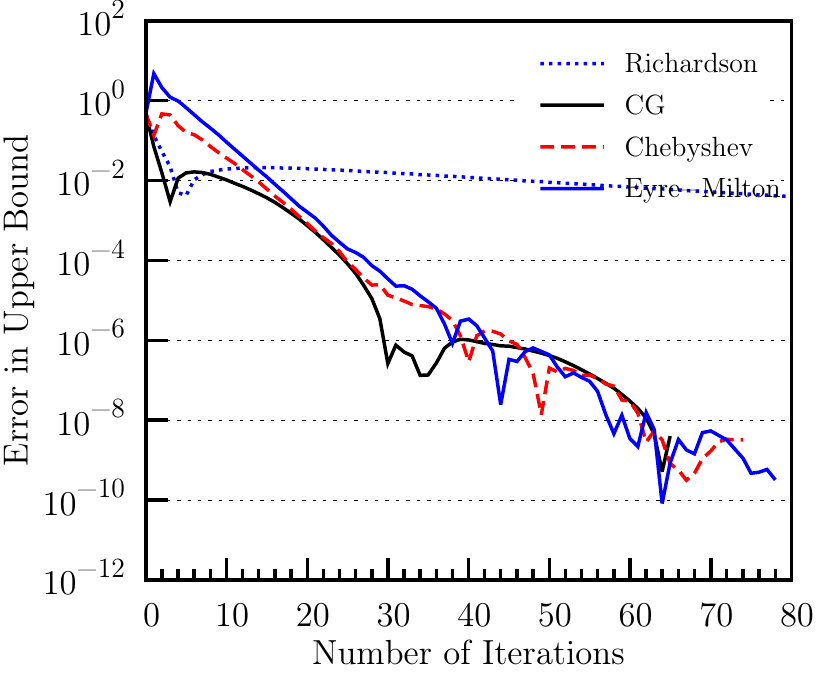}
&
\sublabel{d}\includegraphics[height=55mm]{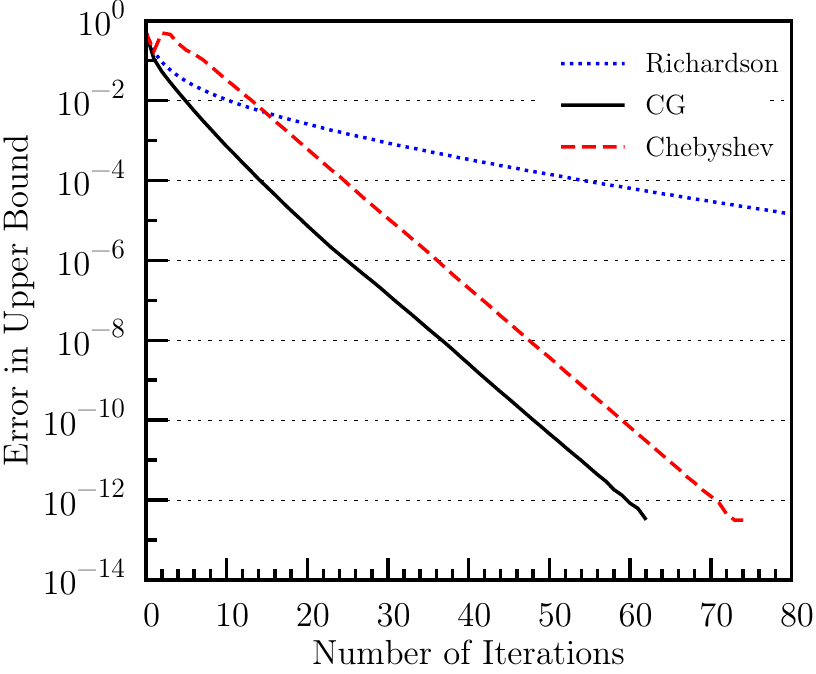}
\end{tabular}
\caption{Convergence of error in guaranteed upper bound for GaNi~(left) and
Ga~(right) discretizations.}
\label{fig:error_UB_convergence}
\end{figure}

In order to investigate the quality of the upper bound in more detail, we introduce the \emph{error} in the guaranteed upper bound at the $m$-th iteration
\begin{align*}
\left| 
U_{\VE}( \MBx_m ) - U_{\VE}( \MBx^* )
\right| 
\text{ for }
m \in \xN_0
\end{align*}
where $\MBx^*$ denotes the approximation to the solution of~\eqref{eq:linsy}
obtained with the Conjugate gradient method and tolerance $\epsilon = 10^{-12}$.
The results are collected in Figure~\ref{fig:error_UB_convergence} and
demonstrate that to achieve a target accuracy of~$10^{-4}$, for instance, the
Conjugate gradient algorithm needs less than $25$ iterations for GaNi
discretization and about $15$~iterations for Ga, whereas the Chebyshev and Eyre-Milton methods require about $10$~additional iterations to reach the same accuracy~(except for fly ash foam microstructure and GaNi discretization, where the methods perform similarly). The iterates generated by the Richardson scheme deliver an accuracy of $10^{-4}$ only for Ga discretization and $70$ iterations, Figure~\ref{fig:error_UB_convergence}(b), or $60$ iterations, Figure~\ref{fig:error_UB_convergence}(d). The superior performance of the Conjugate gradient method in the Ga setting is not surprising; the upper
bound~\eqref{eq:guar_upper_bound_m} is exactly the energy norm that Conjugate
gradients minimize over the Krylov subspaces~\eqref{eq:Krylov}, as has
been recently pointed out by Vond\v{r}ejc~\cite{Vondrejc2015FFTimproved}. 

\begin{figure}[h]
\centering
\begin{tabular}{cc}
\sublabel{a}\includegraphics[height=55mm]{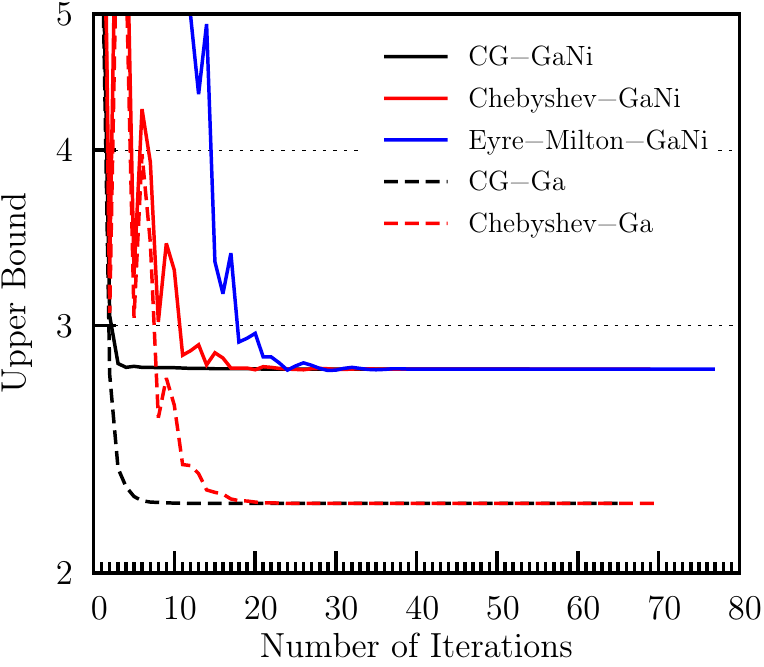}
&
\sublabel{b}\includegraphics[height=55mm]{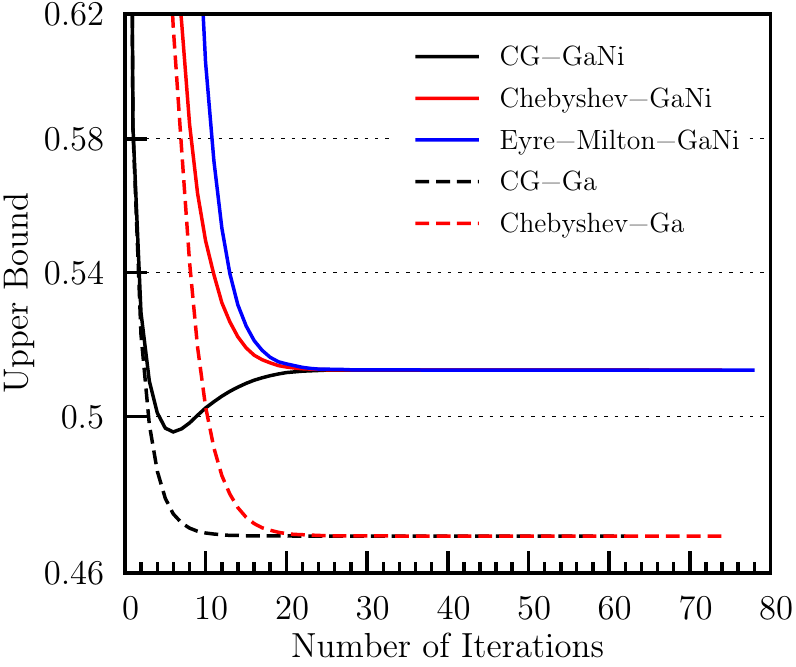}
\end{tabular}
\caption{Effect of numerical integration on tightness of guaranteed
upper bound for (a)~square inclusion and (b)~fly ash foam
microstructures.}
\label{fig:guar_bound_tight}
\end{figure}

Being inspired by recent results in~\cite{Vondrejc2015FFTimproved}, we also find it
instructive to demonstrate the effect of exact numerical integration on the
tightness of the guaranteed upper bounds. Results in
Figure~\ref{fig:guar_bound_tight} confirm that these effects are indeed
significant. For square inclusion and GaNi, the bound equals to $2.793$ and decreases by $20\%$ to $2.241$ by the exact integration; for the fly ash foam
cell, GaNi provides a value of $0.513$ that is improved to $0.469$ by Ga. We
believe that these results promote Ga discretization over GaNi, despite the
increased computational costs due to involvement of a double grid.

\subsection{Non-conformity}\label{sec:incompatiblity}

\begin{figure}[h]
\centering
\begin{tabular}{cc}
\sublabel{a}
\includegraphics[height=55mm]{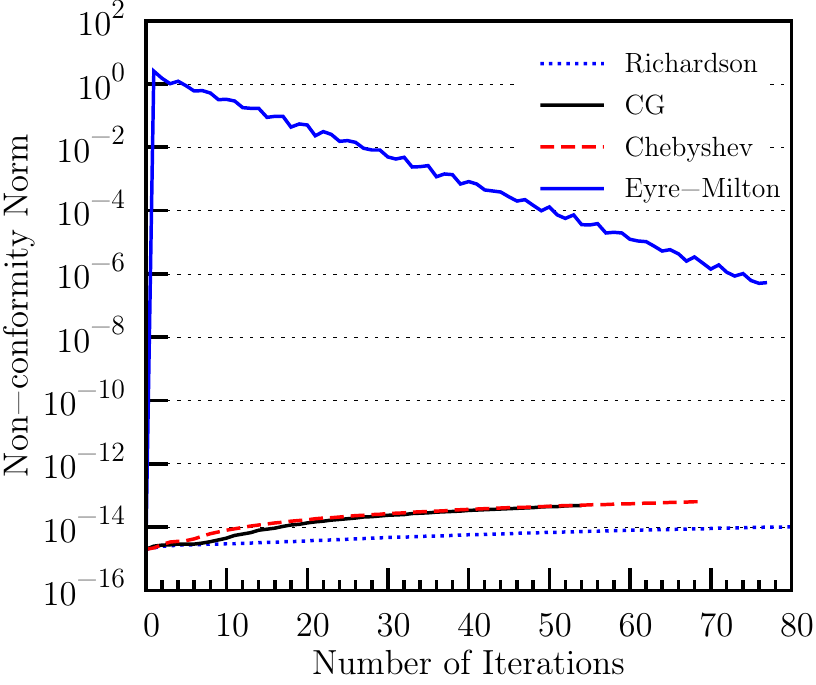}
&
\sublabel{b}
\includegraphics[height=55mm]{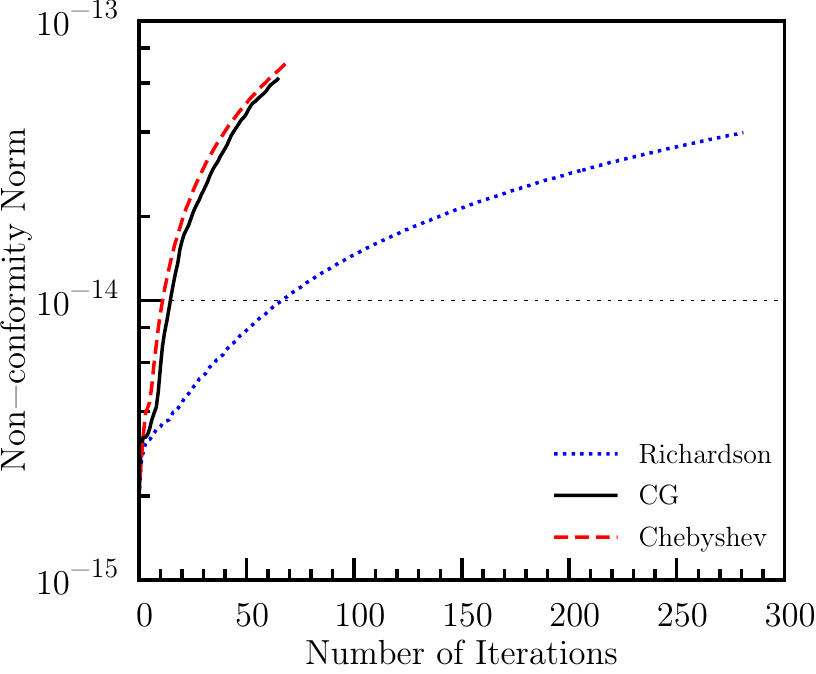}
\end{tabular}
\caption{Evolution of non-conformity norm during iterations for square inclusion and (a)~GaNi and (b)~Ga discretizations.}
\label{fig:incompatibility}
\end{figure}

We shall quantify the non-conformity of iterates for GaNi-based systems by the quantities
\begin{align*}
\norm{\MBGN{}\MBx_m + \MBE_{\VN} - \MBx_m}{\xRdN}, &&
\norm{\MBGN{}\MBx_m-\MBx_m}{\xRdN}
\text{ for } m \in \xN_0,
\end{align*}
defined for the Eyre-Milton scheme and the remaining algorithms, respectively, with an obvious generalization to the Ga setting. The reason for this distinction is best seen in Figure~\ref{fig:incompatibility}(a), which highlights the difference
between the non-conforming Eyre-Milton scheme and conforming solvers -- the Richardson, the Conjugate gradient, and the Chebyshev algorithms. For the Eyre-Milton scheme, the non-conformity error reaches its maximum in the first iteration and progressively decreases in a non-monotone way. For conforming algorithms, the error increases with an increasing number of iterations due to the accumulation of round-off errors. Notice that although the round-off effects per iteration are smallest for the Richardson scheme, the total values are the same at convergence, Figure~\ref{fig:incompatibility}(b).

\begin{figure}
\centering
\includegraphics[height=55mm]{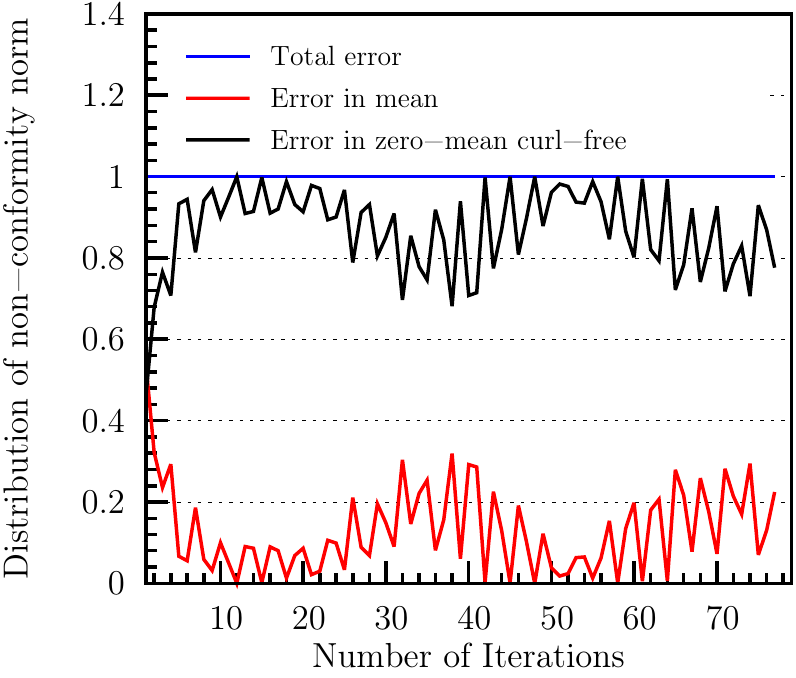}
\caption{Relative distribution of non-conformity norm in the Eyre-Milton algorithm for the square inclusion problem and GaNi discretization.} 
\label{fig:relative_error_EM} 
\end{figure}

As follows from the discrete Helmholtz decomposition of $\xRdN$~\cite[Lemma~21]{VoZeMa2014GarBounds}, the non-conformity error of the Eyre-Milton scheme can be further split into two orthogonal components,
\begin{align*}
\norm{\MBGN{}\MBx_m + \MBE_{\VN} - \MBx_m}{\xRdN}
=
\norm{\MBE_{\VN} - \mean{\MBx_m}}{\xRdN}
+
\norm{\MBG_\VN\MBx_m-\MBx_m + \mean{\MBx_m}}{\xRdN},
\end{align*}
which quantify the difference between the \emph{mean value} of the $m$-iterate, defined with $\mean{\MBx_m^\Vn} =  ( \sum_{\Vk \in \ZNd}{\MB{x}^{\Vk}_m} ) / |\VN|$ for $\Vn \in \ZNd$, and the prescribed vector $\MBE_\VN$, and the distance from \emph{zero-mean curl-free} vectors $\xEN$. The relative distribution of both components, Figure~\ref{fig:relative_error_EM}, demonstrates that the latter component dominates the error in mean.

\subsection{Relative error bound}

\begin{figure}[h]
\centering
\begin{tabular}{cc}
\sublabel{a}\includegraphics[height=55mm]{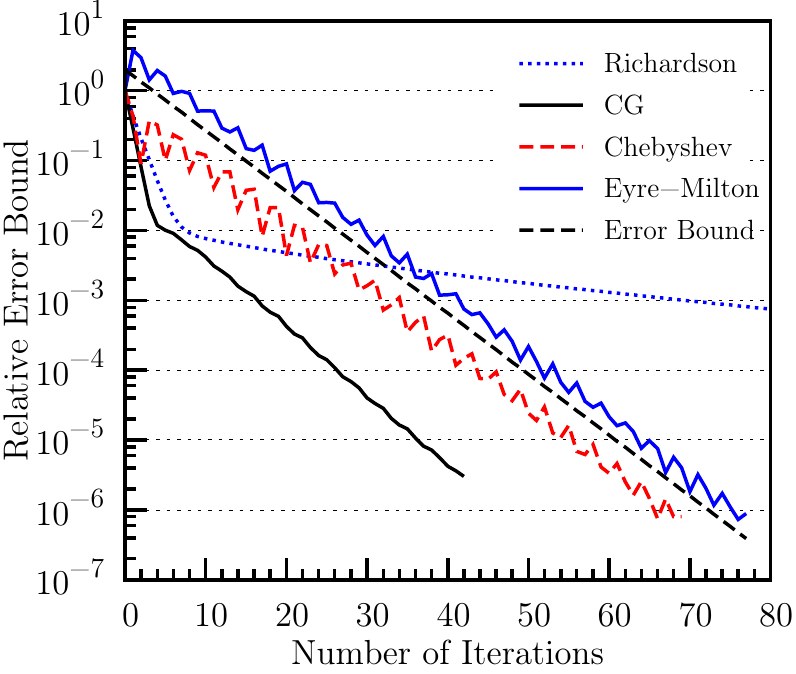} &
\sublabel{b}\includegraphics[height=55mm]{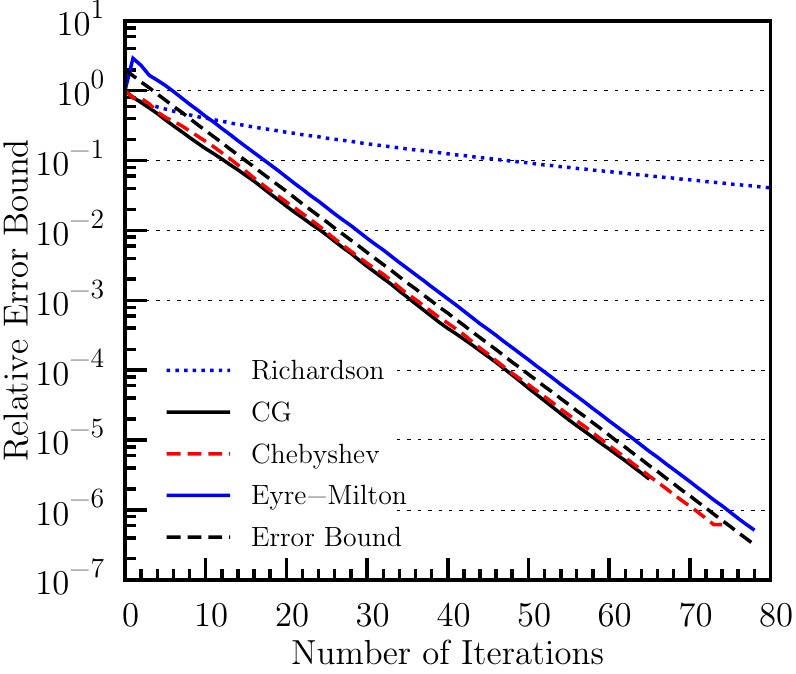}
\end{tabular}
\caption{Performance of error estimates for GaNi discretization and (a)~square
inclusion and (b)~fly ash foam microstructures; error bound is defined by~\eqref{eq:CG_estimate}.}
\label{fig:relative_true_error}
\end{figure}

In support of the theoretical results gathered in
Section~\ref{sec:Lin_solver}, our purpose is to illustrate
the behavior of the relative error
\begin{align*}
\frac{\norm{\MBx_m - \MBx^*}{\MB{C}}}{\norm{\MBx_0-\MBx^*}{\MB{C}}}
\text{ for }
m \in \xN_0.
\end{align*}
The results of this study appear in Figure~\ref{fig:relative_true_error} for the
GaNi-based system and confirm the linear convergence of the Richardson scheme implied by error estimate~\eqref{eq:Richard_estimate} and the
super-linear convergence of the remaining three algorithms, in agreement with
relations~\eqref{eq:CG_estimate} and~\eqref{eq:E-M_estimate}. Notice that no results for Ga discretizations have been shown, since they are very similar to the error plots for the guaranteed upper bound, Figure~\ref{fig:error_UB_convergence}(b,d), and lead to the same conclusions. 

\subsection{Condition number}\label{sec:spectral-radius}

\begin{figure}[h]
\centering
\begin{tabular}{cc}
\sublabel{a}\includegraphics[height=55mm]{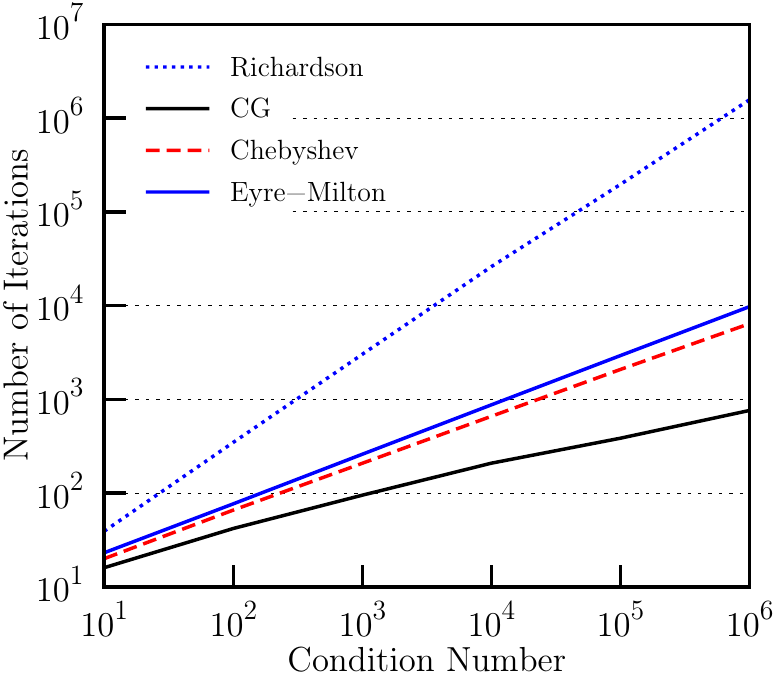}
&
\sublabel{b}\includegraphics[height=55mm]{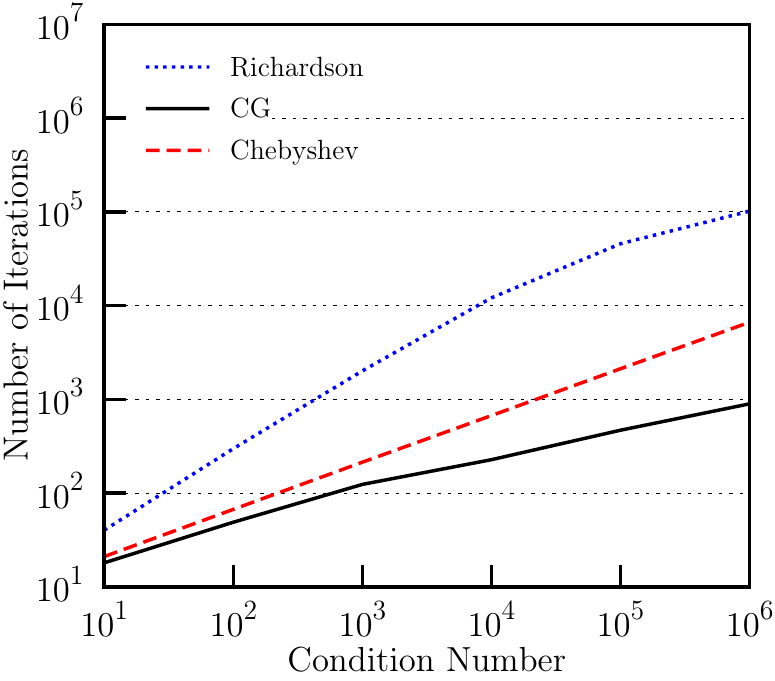}
\end{tabular}
\caption{Dependence of the number of iterations to convergence on the condition number of the system matrix for square inclusion and (a)~GaNi and (b)~Ga discretizations.}
\label{fig:scaling}
\end{figure}

To conclude our numerical experiments, in Figure~\ref{fig:scaling} we plot the
dependence of the number of iterations to reach the accuracy of $\epsilon =
10^{-6}$ on the condition number of system matrix $\kappa$. Note that we present
the results only for the square inclusion, Figure~\ref{fig:uc_types}(a), and set the coefficient of the inclusion equal to $\kappa$~(analogous conclusions hold for more complex microstructures).  In addition, all algorithms are terminated by the relative residual norm.

The results generally follow the trend predicted by the error estimates~\eqref{eq:Richard_estimate}, \eqref{eq:CG_estimate}, and~\eqref{eq:E-M_estimate} and 
confirm the linear scaling $\mathcal{O}(\kappa)$ for the Richardson scheme and
$\mathcal{O}(\sqrt{\kappa})$ for the remaining solvers. Closer inspection
reveals that some solvers converge \emph{faster} for large values of $\kappa$,
namely the Conjugate gradient method or the Richardson scheme for Ga discretization. Such superconvergent behavior was reported in earlier studies by, e.g., Monchiet and
Bonnet~\cite{Monchiet2012polarization} and Willot~et
al.~\cite{Willot2013fourier} for FFT-based solvers, and Schneider~et al.~\cite{Schneider:2016:CHE} for finite difference methods, in order to demonstrate that their methods work for $\kappa \rightarrow \infty$. On the basis of the presented results, one may thus conjecture that iterative methods applied to Ga-based systems will converge for the infinite contrast of material coefficients. We would like to address this question in our future work, together with a more detailed investigation into the distribution of discretization errors.

\section{Conclusions}\label{sec:conclusion}

In this paper, we have performed a comparative study of iterative algorithms for
systems of linear equations arising from a Fourier-Galerkin discretization of
the periodic corrector problem. Two discretization schemes have been considered, 
exact~(Ga) and trapezoidal~(GaNi) integration, and the ensuing systems of linear
equations have been solved with the Richardson, the Chebyshev, the Conjugate gradient,
and the Eyre-Milton algorithms. 

\begin{table}[h]
\centering
\begin{tabular}{lccccc}
\hline
\emph{Algorithm} & \emph{Scaling} & \emph{Conforming} & \emph{Ga} &
\emph{Upper bound} & \emph{Storage} \\
\hline
Richardson & 
$\mathcal{O}(\kappa^1)$ & Yes &  Yes & Monotone & 3 \\
Conjugate gradients &  
$\mathcal{O}(\kappa^\frac{1}{2})$ & Yes & Yes & Monotone & 4 \\
Chebyshev &  
$\mathcal{O}(\kappa^\frac{1}{2})$ & Yes & Yes & Non-monotone & 3 \\ 
Eyre-Milton & 
$\mathcal{O}(\kappa^\frac{1}{2})$ & No & No & Non-monotone & 3 \\
\hline
\end{tabular}
\caption{Overall comparison of the four short-recurrence iterative algorithms;
\emph{scaling} indicates how the number of iterations to converge grows with
an increasing condition number of system matrix $\kappa$, \emph{conforming}
algorithms generate iterates from subspace $\xE$, Ga refers to the extendibility
of the algorithm to the exact integration, \emph{upper bound} indicates 
how the approximate upper bounds generated by the iterates converge to the
guaranteed upper bound, \emph{storage} specifies the number of vectors needed
in one iteration.}
\label{tab:performance_comparison}
\end{table}

Based on the outcomes of our study, summarized in
Table~\ref{tab:performance_comparison} for the reader's convenience, we conclude
that:
\begin{enumerate}
  \item In terms of the \emph{rate of convergence}, the Conjugate gradient, the
  Chebyshev, and the Eyre-Milton algorithms exhibit super-linear
  convergence and the Richardson method converges with the linear rate. In
  addition, the Conjugate gradient method appears to be the most efficient
  solver, while the Chebyshev and the Eyre-Milton algorithms display comparable performance. 
  \item All three general-purpose solvers --- the Richardson, the Conjugate
  gradient, and the Chebyshev algorithms --- generate iterates that \emph{conform} to $\xE$, the space associated with curl-free and zero-mean
  trigonometric polynomials. The Eyre-Milton method produces non-conforming
  iterates.
  \item The general-purpose solvers work for linear systems arising from both Ga and GaNi discretizations, while the Eyre-Milton method is applicable exclusively to the GaNi setting. 
  \item The approximate upper bounds generated by the Richardson and the
  Conjugate gradient methods for the Ga discretizations exhibit \emph{monotone
  convergence}, whereas all other options yield non-monotone convergence. The
  superior performance of the Conjugate gradient algorithm follows from the fact
  that the bound corresponds to the energy norm that Conjugate gradients
  naturally minimize. 
  \item With regard to \emph{memory efficiency} of the implementations introduced in Section~\ref{sec:Lin_solver}, the Conjugate gradients need to store one additional vector per iteration. The computational complexity of a single iteration of all algorithms is comparable because it is dominated by the forward and inverse FFTs.   
\end{enumerate}

We believe that the computational observations collected in this paper provide a
convenient starting point for the development of even more efficient solvers for
FFT-based homogenization algorithms. Our particular interest is to continue to
explore the potential of the Chebyshev method to achieve a more robust
convergence, e.g.~\cite{wathen:2008:cheb}, or to serve as a preconditioner in multi-grid
solvers \cite{parallel:cheby}. Investigations of these topics are underway and
results will be reported separately.

\section*{Acknowledgments}
This work was supported by the European Social Fund within the framework of realizing the project ``Support of inter-sectoral mobility and quality enhancement of research teams at Czech Technical University in Prague'', CZ.1.07/2.3.00/30.0034~(Nachiketa Mishra, Jan Zeman). In addition, Jaroslav Vond\v{r}ejc acknowledges support by the Czech Science Foundation, project No.~13-22230S. Nachiketa Mishra would also like to thank Professor Andrew Wathen~(Oxford University), Dr. Petr Tich\'{y}~(Czech Academy of Sciences), and Professor Zden\v{e}k Strako\v{s}~(Charles University in Prague) for their inspiring comments on many aspects of the results presented in this work. Finally, we would like to thank Stephanie Krueger~(National Library of Technology) for her helpful comments on the manuscript.

\end{document}